\documentclass[10pt,twocolumn,letterpaper]{article}

\usepackage{wacv}              
\usepackage[T1]{fontenc}

\usepackage{ifthen}
\newboolean{comments}
\setboolean{comments}{true} 

\newcommand{\rg}[1]{}
\newcommand{\ag}[1]{}
\newcommand{\ia}[1]{}
\newcommand{\pd}[1]{}
\newcommand{\sh}[1]{}
\newcommand{\ts}[1]{}

\ifthenelse{\boolean{comments}}{

  \setlength{\marginparwidth}{1.2cm}
  \setlength{\marginparsep}{2pt}
  \reversemarginpar

  \renewcommand{\rg}[1]{\textcolor{blue}{RG: #1}}
  \renewcommand{\ag}[1]{\textcolor{blue}{AG: #1}}
  \renewcommand{\ia}[1]{\textcolor{blue}{IA: #1}}
  \renewcommand{\pd}[1]{\textcolor{blue}{PD: #1}}
  \renewcommand{\sh}[1]{\textcolor{blue}{SS: #1}}
  \renewcommand{\ts}[1]{\textcolor{blue}{[TS: #1]}}

}{}

\definecolor{wacvblue}{rgb}{0.21,0.49,0.74}
\usepackage[pagebackref,breaklinks,colorlinks,allcolors=wacvblue]{hyperref}

\def\wacvPaperID{934} 
\def\confName{WACV}
\def\confYear{2027}

\title{FASHION-RQ: Mining Hierarchical Semantic Structure for Automated Evaluation of Fashion Retrieval}

\author{
Tripti Shukla\\
Glance\\
Bengaluru, India\\
{\tt\small tripti.shukla@glance.com}
\and
Shwetha Somasundaram\\
Glance\\
Bengaluru, India\\
{\tt\small shwetha.s@glance.com}
\and
Akshat Gupta\\
Glance\\
Bengaluru, India\\
{\tt\small akshat.gupta@glance.com}
\and
Rohit Garg\\
Glance\\
Bengaluru, India\\
{\tt\small rohit.garg@glance.com}
\and
Paul Duff\\
Glance\\
Vilnius, Lithuania\\
{\tt\small paul.duff@glance.com}
\and
Ian Anderson\\
Glance\\
London, United Kingdom\\
{\tt\small ian.anderson@glance.com}
}

\begin{document}
\maketitle
\begin{abstract}
    
Evaluating fashion retrieval systems remains challenging due to the limited
availability of benchmarks that capture the fine-grained, open-ended queries
encountered in real-world search and recommendation. Existing benchmarks
often rely on manual curation, predefined taxonomies, or synthetic prompts,
limiting scalability and semantic coverage. We introduce
\textbf{FASHION-RQ: Mining Hierarchical Semantic Structure for Automated
Evaluation of Fashion Retrieval}, an automated framework that derives
user-type queries and relevance sets directly from catalog structure. Our
key insight is that hierarchical discrete representations learned by
Residual Quantized Variational Autoencoders (RQ-VAE) organize products into
semantically coherent clusters at progressively finer granularities. We
leverage this structure to construct \textbf{FASHION-RQBENCH}, a scalable
benchmark without manual query annotation or predefined taxonomies. We
propose four complementary tasks for evaluating ranking, relevance
classification, fine-grained sibling discrimination, and paraphrase
robustness, alongside query- and attribute-level relevance metrics.
Experiments across recent retrieval models demonstrate that
FASHION-RQBENCH differentiates model capabilities across query types and
attributes. A human study further shows that benchmark queries achieve
comparable naturalness and higher intent clarity than real user queries,
supporting their validity for real-world evaluation. The code and
FASHION-RQBENCH dataset will be publicly released upon acceptance.
\end{abstract}

\begin{figure*}[]
  \centering
 \includegraphics[width=\linewidth]{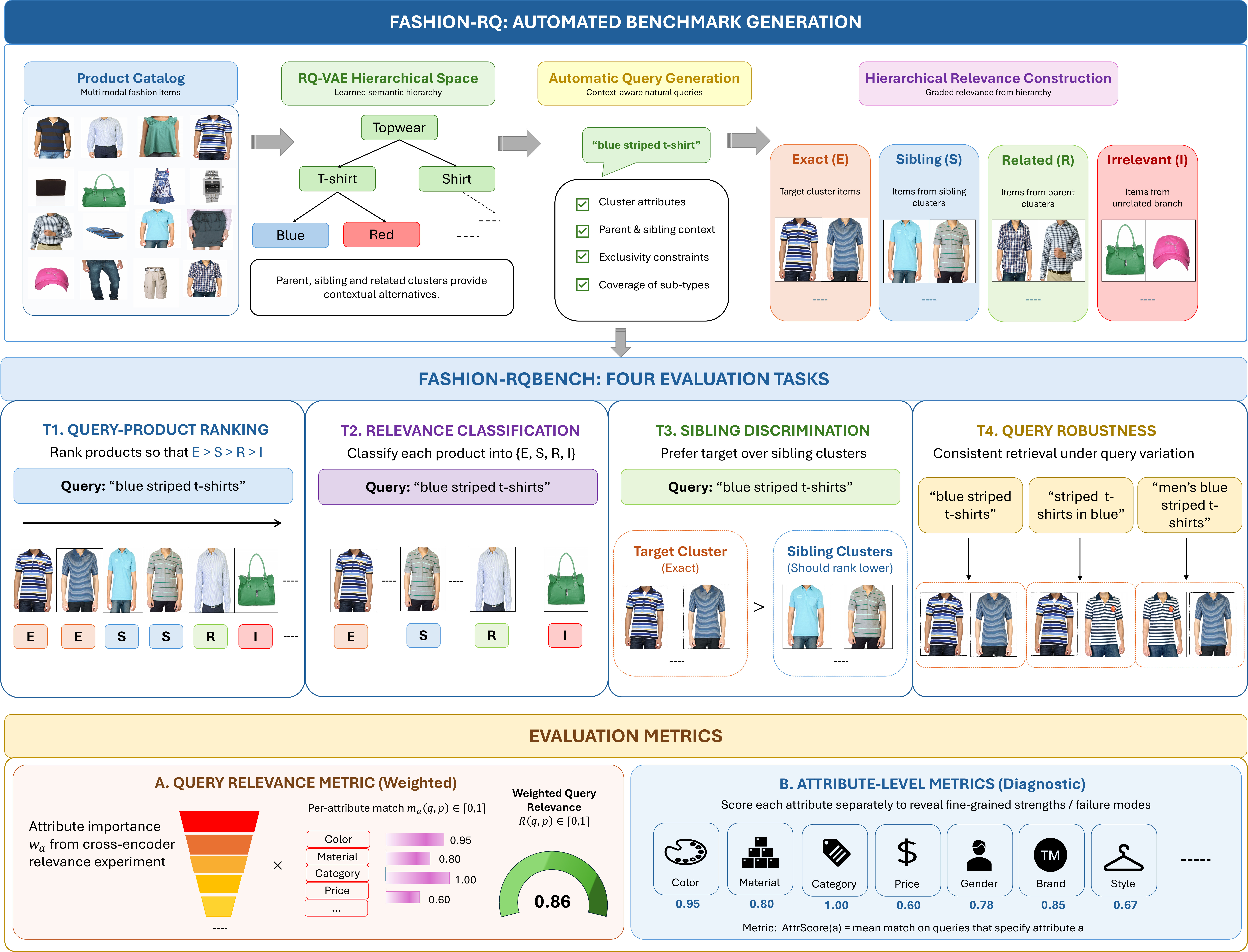}
  \caption{Overview of FASHION-RQ. The framework leverages an RQ-VAE hierarchy to automatically generate natural-language queries and graded relevance sets, with Exact, Sibling, Related, and Irrelevant relationships. The resulting benchmark supports four complementary retrieval tasks, while weighted query relevance and attribute-level metrics evaluate overall intent satisfaction and fine-grained attribute matching.}
  \vspace{-5mm}
  \label{fig:overview}
  
\end{figure*}
\section{Introduction}
\label{sec:intro}

Fashion retrieval is inherently \emph{set-valued} and \emph{graded}.
A query such as ``lightweight cotton kurtas for summer'' constrains
multiple attributes, including product type, material, and season, and
many catalog items may satisfy these constraints to different degrees.
Relevance is therefore compositional: an item may match some attributes
while violating others. A single aggregate ranking score can hide these
differences. For example, two systems may achieve similar nDCG while one
retrieves the correct garment in the wrong color and the other retrieves
the correct color for the wrong garment. Evaluating such systems requires
relevance labels that are set-valued, graded, and decomposable by
attribute.

Existing fashion benchmarks do not provide these properties. Manual
fashion retrieval datasets typically associate a query with a single
target item~\cite{liu2016deepfashion,wu2021fashioniq}, penalizing other
valid results. Graded set-valued annotations are possible, but expensive:
ESCI contains 2.6M judgments for 130K queries~\cite{reddy2022shopping}.
Moreover, manually collected labels are tied to a particular catalog
snapshot and query distribution. Automatic assessment using language
models~\cite{upadhyay2024umbrela} removes annotation cost, but typically
produces an independent relevance score for each query--item pair and
does not identify which query attributes are satisfied or violated.
These approaches also treat items independently, overlooking the
relationships among products in the catalog.

We instead observe that the catalog itself contains the structure needed
to define relevance sets. Recent retrieval and recommendation systems such
as DSI~\cite{tay2022dsi}, RIPOR~\cite{zeng2024ripor}, and
TIGER~\cite{rajput2023tiger} represent items using sequences of discrete
codes learned with residual quantization~\cite{lee2022autoregressive}.
Because these codes are organized from coarse to fine, they induce a
prefix tree over the catalog: items sharing a prefix form increasingly
specific groups. We repurpose this structure not as an index for
retrieval, but as an \emph{automatic annotation mechanism}. However,
vanilla RQ-VAE does not guarantee that a code corresponds to a single
semantic attribute, which can produce heterogeneous clusters. We
therefore introduce an \emph{attribute exclusivity loss} that encourages
each quantization level to specialize to a coherent attribute, making
the resulting hierarchy suitable for defining relevance sets. A prefix
then defines a relevance set, its length controls granularity, and the
shared-prefix length between an item and the target prefix provides
graded relevance without a hand-authored taxonomy.

Building on this hierarchy, we filter clusters that are too small or
heterogeneous to serve as relevance sets and compute each retained
cluster's \emph{relational context}: the attributes that distinguish it
from its parent and sibling clusters. A language model then verbalizes
this structured evidence into shopper-style queries. Importantly, the
language model generates the query but does not determine relevance;
relevance is inherited deterministically from the semantic ID (SID)~\cite{rajput2023tiger} structure.

We instantiate this framework on the public Myntra \emph{Fashion Product
Images} catalog~\cite{param_aggarwal_2019}, containing $44{,}446$ items,
and construct \textbf{FASHION-RQBENCH}: $588$ clusters across three
granularity levels and $5{,}380$ attribute-enriched queries. The benchmark
provides four complementary tasks: query--product ranking, hierarchical
relevance classification, sibling discrimination, and paraphrase
robustness. We additionally introduce a weighted query-relevance score
and per-attribute scores to distinguish overall retrieval quality from
fine-grained attribute satisfaction. Figure~\ref{fig:overview} provides an overview of the complete FASHION-RQBENCH pipeline, from hierarchical benchmark construction to query generation, evaluation tasks, and metrics. 

Evaluation of ten off-the-shelf sparse and dense retrieval models,
without fine-tuning, shows that aggregate ranking performance alone can
mask important weaknesses. Dense encoders achieve substantially higher
ranking quality than BM25~\cite{robertson2009probabilistic} ($\mathrm{nDCG@10}=0.87$ vs.\ $0.77$), yet the
best model achieves only $0.51$ accuracy on four-way hierarchical
relevance classification. Model scale is also not consistently predictive:
the $4$B Qwen3~\cite{qwen3_embedding_2025} encoder slightly outperforms its $8$B counterpart on
sibling discrimination. Attribute-level evaluation further localizes
these differences: models perform strongly on product type and gender
($>0.93$) but substantially worse on material ($\approx0.52$) and style
($\approx0.43$). Finally, a human study comparing benchmark queries with
real user queries shows comparable perceived naturalness and strong
intent clarity, supporting Fashion-RQBENCH as a proxy for real-world
fashion search. Our contributions are:
\begin{itemize}

  \item We repurpose RQ-VAE semantic identifiers as an automatic
  \emph{annotation mechanism} and introduce an attribute exclusivity loss
  that encourages quantization levels to encode coherent attributes.
  Prefixes define relevance sets, prefix length controls granularity, and
  shared-prefix length defines graded relevance.

  \item We introduce a cluster-conditioned query generation procedure
  that captures each cluster's attribute distribution and its relational
  differences from parent and sibling clusters, enabling automatic
  generation of shopper-style queries.

  \item We introduce \textbf{FASHION-RQBENCH}, comprising $588$ clusters,
  $5{,}380$ attribute-enriched queries, and $44{,}446$ products, with
  deterministic graded relevance and four complementary evaluation tasks.

  \item We propose weighted query-relevance and per-attribute metrics that
  distinguish overall intent satisfaction from fine-grained attribute
  satisfaction.
\end{itemize}

\section{Related Works}
\label{sec:rel_works}

\noindent\textbf{Fashion Retrieval Datasets and Benchmarks.}
\label{sec:rel-fashion-data}
Fashion retrieval involves fine-grained attributes such as product type,
color, pattern, material, occasion, season, and price. Existing resources
address different aspects of this problem. DeepFashion~\cite{liu2016deepfashion}
and DeepFashion2~\cite{ge2019deepfashion2} provide large-scale benchmarks
for fashion recognition, detection, segmentation, and retrieval;
Fashion200K~\cite{han2017fashion200k} and Shopping100k~\cite{ak2018shopping100k}
support attribute-based search; FACAD~\cite{yang2020fashioncaptioning}
focuses on fashion captioning; and Fashion Product Images
~\cite{param_aggarwal_2019} provides the catalog schema used in our work.
Language-conditioned retrieval is studied primarily through composed image
retrieval, including FashionIQ~\cite{wu2021fashioniq}, and more general
benchmarks such as CIRR~\cite{liu2021cirr}, CIRCO~\cite{baldrati2023searle},
and LaSCo~\cite{levy2024lasco}, with fashion-oriented encoders including
FashionCLIP~\cite{chia2022fashionclip}, Marqo-FashionSigLIP
~\cite{marqo2024fashionsiglip}, and GCL~\cite{zhu2024gcl}.

Most existing fashion benchmarks use \emph{single-target} relevance,
whereas open-ended product queries can correspond to a set of valid
items; CIRCO~\cite{baldrati2023searle} is a notable exception, but at
modest scale. Graded set-valued relevance is available in product search:
ESCI~\cite{reddy2022esci}, for example, provides exact, substitute,
complement, and irrelevant labels, but requires 2.6M manual judgments for
130K queries. Manual construction also fixes the query distribution and
catalog snapshot, making large-scale regeneration costly. Outside fashion,
QUEST~\cite{malaviya2023quest} automatically constructs set-valued queries
from Wikipedia categories, but relies on an existing taxonomy. In contrast,
FASHION-RQ derives both relevance sets and queries from the catalog's
learned semantic structure, avoiding manual relevance annotation and
predefined taxonomies.

\noindent\textbf{Generative Retrieval and Semantic Identifiers.}
\label{sec:rel-genret}
Generative retrieval directly decodes item identifiers from queries.
DSI~\cite{tay2022dsi} and NCI~\cite{wang2022nci} established this paradigm,
with hierarchical identifier construction shown to be effective.
Subsequent approaches learn identifiers through discrete auto-encoding
~\cite{sun2023genret} or residual quantization for relevance-aware,
prefix-scorable representations~\cite{zeng2024ripor}. In recommendation,
TIGER~\cite{rajput2023tiger} represents items using SIDs produced
by an RQ-VAE~\cite{lee2022autoregressive,oord2017vqvae}; its coarse-to-fine
codebooks induce a prefix hierarchy, while Singh~\etal.\
~\cite{singh2023semanticids} demonstrate benefits for cold-start and
long-tail recommendation.

Recent work explicitly shapes this hierarchy. HiD-VAE~\cite{fang2025hidvae}
supervises quantization levels with hierarchical item tags and introduces
a uniqueness objective to reduce code collisions. These approaches,
however, use semantic identifiers primarily as representations for
downstream retrieval or recommendation. FASHION-RQ instead repurposes the
hierarchy as an \emph{annotation structure}: prefixes define product
groups at different granularities and shared-prefix length provides
graded relevance. We further introduce level-wise attribute supervision
with an \emph{attribute exclusivity loss} to encourage semantically
coherent clusters, using the resulting tree to construct the benchmark
rather than the retrieval identifiers themselves.

\noindent\textbf{Automated Query and Relevance Construction.}
\label{sec:rel-autogen}
Synthetic query generation reduces the cost of manual annotation.
Doc2Query~\cite{nogueira2019doc2query} generates queries for documents,
while InPars~\cite{bonifacio2022inpars} and Promptagator
~\cite{dai2023promptagator} use LLM-generated query--document pairs for
retrieval training. These approaches are primarily
\emph{document-conditioned}: a query is generated from an individual
document and typically inherits a single positive target. Rahmani~\etal.\
~\cite{rahmani2024synthetic} extend synthetic generation to evaluation,
using LLM-generated queries and judgments to reproduce retrieval rankings,
but the resulting relevance labels remain pairwise LLM assessments.
BEIR~\cite{thakur2021beir}, in contrast, aggregates existing manually
annotated datasets, while ExcluIR~\cite{zhang2024excluir} focuses on
phenomenon-specific retrieval with manually constructed queries.

FASHION-RQ uses \emph{cluster-conditioned} generation: queries are
generated from a purified cluster's attribute distribution together with
its relational differences from parent and sibling clusters. The cluster
itself therefore defines a multi-item relevance set, while the relational
context encourages queries that distinguish the target group from nearby
alternatives. The LLM only verbalizes catalog-derived evidence; it does
not assign relevance, keeping the resulting labels independent of the
generator.

\noindent\textbf{Evaluating Retrieval Beyond Binary Relevance.}
\label{sec:rel-eval}
Graded metrics such as nDCG~\cite{jarvelin2002ndcg} require relevance
judgments, motivating LLM-based assessors such as UMBRELA
~\cite{upadhyay2024umbrela}. Such assessors produce holistic query--item
scores and therefore do not reveal which attributes are satisfied or
violated. Attribute-level evaluation has been explored through
attribute-specific retrieval metrics~\cite{ma2020asen}, conditional
similarity~\cite{vaze2023genecis}, compositionality benchmarks
~\cite{yuksekgonul2023aro,hsieh2023sugarcrepe}, and constraint-satisfaction
measures~\cite{abdin2023kitab}. However, these settings either evaluate
individual attributes, use curated distractors, or operate over restricted
constraint types rather than decomposing relevance over a ranked product
catalog.

FASHION-RQ combines set-valued graded relevance with joint
attribute-level evaluation. Our weighted query-relevance score and
per-attribute metrics are computed from the structured query
representation $\phi(q)$, separating overall intent satisfaction from
fine-grained attribute satisfaction. Because relevance is inherited
deterministically from the semantic-ID hierarchy and attribute scores use
a fixed catalog schema, the evaluation is reproducible and inspectable
without per-pair human or LLM judgments.

\section{FASHION-RQ: Automated Benchmark Generation}
\label{sec:method}

We introduce \textbf{FASHION-RQ}, an automated framework for constructing
retrieval benchmarks from product catalogs without manual query or relevance
annotation. It exploits the hierarchical structure induced by residual
quantization: SID prefixes define relevance sets at multiple
granularities, while catalog statistics generate natural-language queries
that represent the target cluster and discriminate it from nearby alternatives.

Given a catalog $\mathcal{I} = \{i_1,\ldots,i_N\}$, each item $i$ is represented by a dense embedding $\mathbf{e}_i \in \mathbb{R}^{1024}$ and catalog metadata including product type, category, color, gender, usage, price, and textual information. A hierarchical tokenizer maps each item to a SID $s_i = (s_i^{(1)},s_i^{(2)},s_i^{(3)})$, with $L=3$ quantization levels and $K=64$ codes per level. A prefix
$p^{(\ell)} = (s^{(1)},\ldots,s^{(\ell)})$ defines the cluster
$C\!\left(p^{(\ell)}\right)
=
\left\{
i \in \mathcal{I} :
s_i^{(1:\ell)} = p^{(\ell)}
\right\}$,
where $\ell=1$ and $\ell=3$ correspond to the coarsest and finest
granularities, respectively.

Our benchmark construction consists of four stages:
(1) \emph{hierarchical semantic representation}, which induces a
coarse-to-fine catalog hierarchy;
(2) \emph{cluster purification}, which removes unreliable clusters;
(3) \emph{relational context construction}, which identifies attributes
distinguishing each cluster from its parent and siblings; and
(4) \emph{automatic query generation}, which produces queries satisfying
coverage and discriminability requirements.
The resulting benchmark contains natural-language queries paired with
automatically derived relevance sets at multiple hierarchy levels.

\noindent \textbf{Problem Formulation.} \label{sec:problem}
Fashion retrieval should be evaluated across multiple levels of semantic
granularity, as queries may target broad product groups or fine-grained
subsets. FASHION-RQ uses semantic IDs (SIDs) to construct such an evaluation
hierarchy: each SID prefix $p^{(\ell)}$ defines a catalog cluster
$C(p^{(\ell)})$, from which we generate natural-language queries
$Q(p^{(\ell)})$ and derive their ground-truth relevance sets.

We use an RQ-VAE to induce this hierarchy, as residual quantization captures
entangled attribute interactions and organizes products into semantically
coherent clusters at multiple granularities. This structure provides a
catalog-grounded basis for realistic query generation and reproducible
evaluation of fashion retrieval systems.

With $K=64$ codes and $L=3$ levels, the hierarchy admits
$\sum_{\ell=1}^{L}K^\ell=64+64^2+64^3\approx2.7\times10^5$ prefixes,
making manual benchmark construction impractical. FASHION-RQ therefore
automatically selects prefixes, derives their relevance sets, and generates
queries:
\begin{equation}
\mathcal{B}=
\left\{
(q_j,p_j^{(\ell)},C(p_j^{(\ell)}),\phi(q_j))
\right\}_{j=1}^{N},
\end{equation}
where $q_j$ is the natural-language query, $p_j^{(\ell)}$ its semantic
granularity, $C(p_j^{(\ell)})$ its relevance set, and $\phi(q_j)$ its
structured attributes.

\begin{figure}[h]
  \centering
 \includegraphics[width=\linewidth]{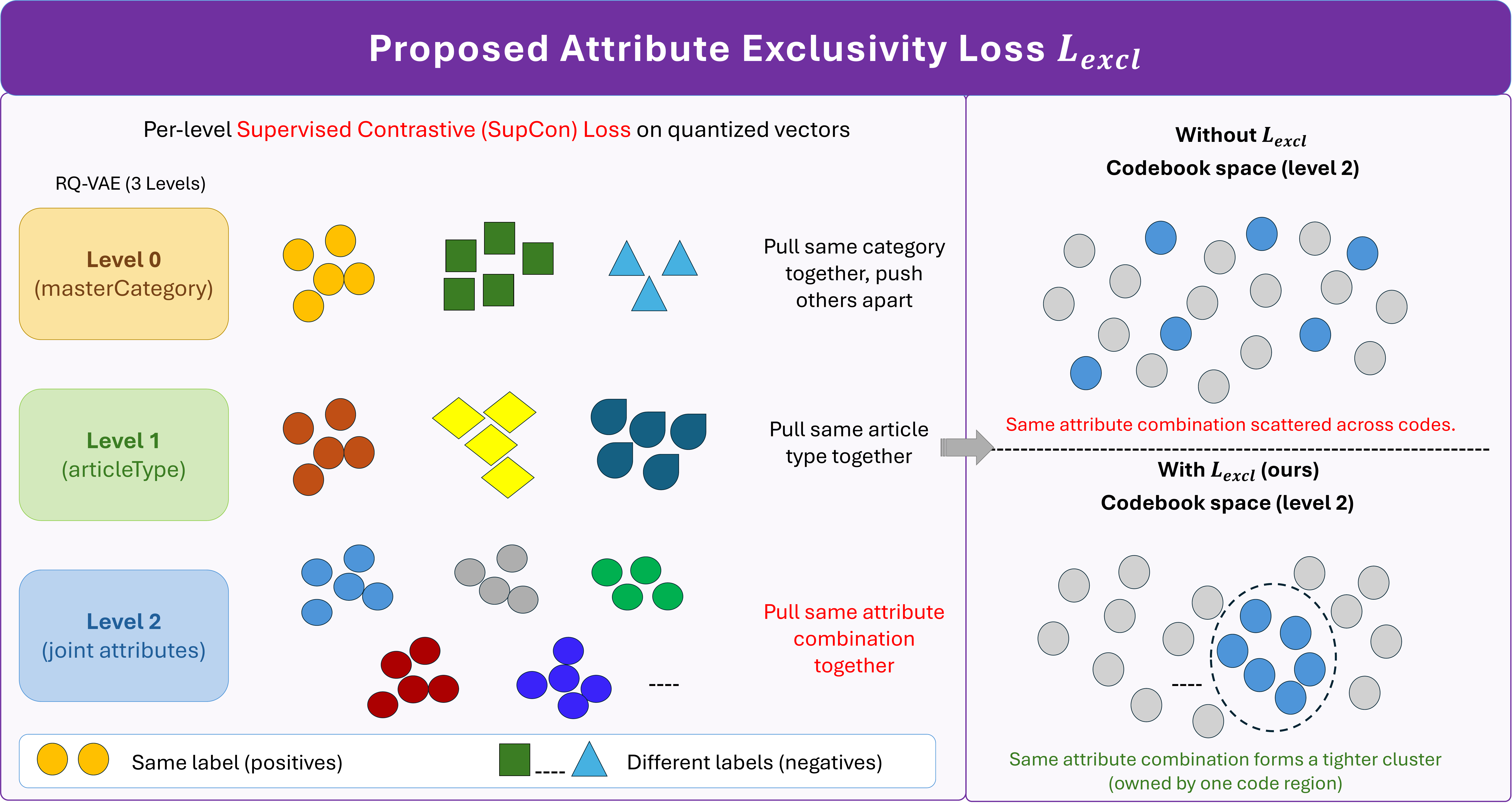}
  \caption{Attribute Exclusivity Loss. Per-level supervised contrastive supervision encourages progressively finer attribute semantics to form compact, distinct regions in the RQ-VAE codebooks, reducing the scattering of identical attribute combinations across codes.}
  \label{fig:excl_loss}
\end{figure}

\noindent \textbf{Hierarchical Semantic Representation.}
\label{sec:hier-rep} We first represent each product using a multimodal embedding from
Marqo-FashionSigLIP~\cite{marqo2024fashionsiglip}, trained with Generalized
Contrastive Learning (GCL) on rich product signals, including product images, descriptions,
categories, style, color, material, keywords, and fine-grained details,
across over one million fashion products. This provides a semantically rich
representation beyond text-only descriptions.

Given $\mathbf{e}_i\in\mathbb{R}^{1024}$, an MLP encoder $f_\theta$ maps each product to $z_i=f_\theta(\mathbf{e}_i)\in\mathbb{R}^{32}$
with architecture $1024\!\rightarrow\!512\!\rightarrow\!256\!\rightarrow\!128 \!\rightarrow\!32$. We apply residual quantization~\cite{lee2022autoregressive}
with $L=3$ codebooks of $K=64$ codes. 
Starting from $r_i^{(0)}=z_i$, residual quantization selects the code and
updates the residual at each level:
\begin{equation}
\begin{aligned}
s_i^{(\ell)}
&=
\arg\min_{k\in\{1,\ldots,K\}}
\left\|r_i^{(\ell-1)}-\mathbf{c}_k^{(\ell)}\right\|_2^2,\\
r_i^{(\ell)}
&=
r_i^{(\ell-1)}-\mathbf{c}_{s_i^{(\ell)}}^{(\ell)}.
\end{aligned}
\end{equation}
The quantized representation is
$\hat{z}_i=\sum_{\ell=1}^{L}\mathbf{c}_{s_i^{(\ell)}}^{(\ell)}$.
Thus, the hierarchy captures coarse-to-fine product structure.

For benchmark construction, the induced clusters should be both
\emph{exhaustive} and \emph{exclusive}: items sharing an attribute
combination should be grouped together, while distinct combinations should
map to distinct codes. Reconstruction and VQ losses promote exhaustivity by
preserving the input embedding, but do not explicitly prevent semantically
different combinations from sharing a code. We therefore augment the
objective with an attribute-based exclusivity loss:
\begin{equation}
\mathcal{L}
=
\mathcal{L}_{\mathrm{recon}}
+
\sum_{\ell=1}^{L}
\left(
\mathcal{L}_{\mathrm{codebook}}^{(\ell)}
+
\beta\mathcal{L}_{\mathrm{commit}}^{(\ell)}
\right)
+
\lambda_{\mathrm{excl}}\mathcal{L}_{\mathrm{excl}}.
\end{equation}

The exclusivity loss applies supervised contrastive learning~\cite{khosla2021supervisedcontrastivelearning} directly to
the quantized code vectors
$q_i^{(\ell)}=\mathbf{c}_{s_i^{(\ell)}}^{(\ell)}$, keeping the supervision
local to each codebook. Figure~\ref{fig:excl_loss} illustrates this level-wise exclusivity supervision. We use \textit{masterCategory} at level~1,
\textit{articleType} at level~2, and all articleType-anchored joint
attributes \textit{(color, gender, usage, season, age group, price tier, and fashion
type)} at level~3. For attributes with at least one valid positive pair in a
mini-batch, the corresponding losses are averaged:
\begin{equation}
\mathcal{L}_{\mathrm{excl}}
=
\sum_{\ell=1}^{L}
\frac{1}{|\mathcal{A}_{\mathrm{valid}}^{(\ell)}|}
\sum_{a\in\mathcal{A}_{\mathrm{valid}}^{(\ell)}}
\operatorname{SupCon}(q^{(\ell)},y_a;\tau).
\end{equation}
This prevents sparse attributes from diluting the available supervision
while enforcing exclusivity directly at each discrete level. We set $\tau=0.07$, following supervised contrastive learning
~\cite{khosla2021supervisedcontrastivelearning}, and
$\lambda_{\mathrm{excl}}=0.05$ to keep the exclusivity term subordinate to the primary reconstruction and quantization losses. 
At inference, SIDs are obtained
deterministically using the same residual quantization procedure, yielding
$s_i\in\{0,\ldots,63\}^{3}$.

\noindent\textbf{Cluster Purification.}
\label{sec:purification}
The induced hierarchy may contain small or heterogeneous clusters.
We therefore purify it post-hoc, independent of RQ-VAE training.
Using \textit{articleType} as the anchor attribute, cluster purity is
the fraction of labeled items belonging to the dominant product type:
\begin{equation}
\operatorname{Purity}(p^{(\ell)})=
\frac{\max_a|\{i\in C(p^{(\ell)}):\texttt{aT}_i=a\}|}
{|\{i\in C(p^{(\ell)}):\texttt{aT}_i\neq\texttt{null}\}|}.
\end{equation}
We retain prefixes with $|C(p^{(\ell)})|\geq5$ and
$\operatorname{Purity}(p^{(\ell)})\geq0.80$; at the finest level, we
additionally require at least two items. All qualifying prefixes are
retained to preserve hierarchy coverage.

\noindent\textbf{Relational Context.}
\label{sec:relational-context}
Purity does not capture how a cluster differs from nearby alternatives.
For each surviving prefix $p^{(\ell)}$, we compare it with its parent and
up to three siblings. For categorical attribute value $(a,v)$, we compute
\begin{equation}
\Delta_{a=v}(p^{(\ell)},R)
=
\operatorname{pct}_{C(p^{(\ell)})}(a=v)-\operatorname{pct}_{R}(a=v),
\end{equation}
where $R$ denotes the parent or a sibling. For numeric attributes, we
compare mean, median, minimum, maximum, and quartiles. Differences of at
least 20 percentage points are retained as exclusivity constraints,
identifying attributes that distinguish the target from nearby clusters.

\noindent\textbf{Automatic Query Generation.}
\label{sec:query-gen}
For each purified cluster, we generate $Q(p^{(\ell)})$ from its attribute
distribution and relational context, targeting \emph{exhaustiveness} and
\emph{exclusivity}. Each sub-type occupying $\geq5\%$ of the cluster receives
at least one query; a catch-all query covers the remainder. Attribute
combinations covering $\geq90\%$ \& $\geq50\%$ of a cluster define
discriminative and coverage signatures, respectively.

We use an instruction-following LLM (Gemini~\cite{geminiteam2025geminifamilyhighlycapable}; temperature $0.2$, top-$p$
$0.8$) whose prompt includes these signatures, attribute distributions,
parent/sibling context, and exclusivity constraints. It is instructed to
produce complete shopper-style queries using discriminative attributes,
verify separation from siblings, and avoid unsupported universal claims.
The LLM returns $6$--$12$ queries plus cluster insights, validation checks,
and a coverage map linking major product types to queries. Outputs are
parsed leniently; the full prompt and schema are provided in the
supplementary material.

\noindent\textbf{Query Attribute Enrichment.}
\label{sec:query-enrichment}
For attribute-aware evaluation, each unique generated query is converted
by the same LLM into
$\{product\_type,gender,age\_group,color,\\pattern,material,occasion,
brand,price\_hint,style\_\\tags\}$. Unexpressed attributes are \texttt{null},
multi-valued attributes may contain multiple values, and \textit{brand} is
populated only when explicitly mentioned. Duplicate queries are removed
globally before enrichment.

\section{Benchmark Definition \& Tasks.}
\label{sec:benchmark}
\noindent \textbf{Source Dataset.}
We propose FASHION-RQBENCH, a benchmark for evaluating fine-grained fashion retrieval and ranking. It is constructed from the public Myntra \emph{Fashion Product Images}~\cite{param_aggarwal_2019} catalog, comprising $44{,}446$ products with structured metadata, free-text descriptions, and product images. Each item is represented by a $1024$-dimensional embedding generated using Marqo-FashionSigLIP~\cite{marqo2024fashionsiglip}, and assigned a 3-level SID using the RQ-VAE model described in Section~\ref{sec:hier-rep}. Table~\ref{tab:benchmark_stats} provides an overview of Fashion-RQBENCH. More details are provided in the supplementary.

\begin{table}[h]
\centering
\caption{Overview of the Fashion-RQBENCH evaluation set.}
\label{tab:benchmark_stats}
\resizebox{\linewidth}{!}{
\begin{tabular}{lcc}
\toprule
\textbf{Statistic} & \textbf{Value} & \textbf{Details} \\
\midrule
Master categories & 5 & Apparel, Accessories, Footwear, Personal Care, Free Items \\
Subcategories & 45 & -- \\
Article types & 143 & -- \\
Base colors & 47 & -- \\
Genders & 5 & Men, Women, Unisex, Boys, Girls \\
\midrule
Benchmark clusters & 588 & 17 / 135 / 436 at levels 1 / 2 / 3 \\
Enriched queries & 5,380 & 2--12 per cluster \\
Unique query strings & 5,305 & 98.7\% of queries \\
Mean query length & 10.0 & 3--26 words \\
\midrule
Level-1 cluster size & 823.9 & Range: 6--5,349 \\
Level-2 cluster size & 219.2 & Range: 5--4,896 \\
Level-3 cluster size & 79.5 & Range: 5--4,888 \\
\bottomrule
\end{tabular}}
\end{table}

\noindent \textbf{Generated Benchmark.}
We exploit the hierarchical RQ-VAE SID structure to construct benchmark instances without manual taxonomy design. Each qualifying 1-, 2-, or 3-level SID prefix that satisfies the purity and minimum-size criteria is materialized as a cluster record. This yields $588$ clusters, comprising $17$, $135$, and $436$ clusters at levels 1, 2, and 3, respectively (Table~\ref{tab:benchmark_stats}). Attribute values exhibit substantial lexical diversity, supporting evaluation beyond a fixed vocabulary.

\noindent \textbf{Benchmark Tasks.}
\label{sec:bench_tasks}
We define four complementary tasks over the same SID hierarchy, with candidate construction and relevance labels derived deterministically, eliminating the need for an independently authored product taxonomy. For a level-3 anchor cluster, candidates receive four graded relevance levels: \emph{Exact} (same level-3 prefix), \emph{Sibling} (same level-2 prefix), \emph{Related} (same level-1 prefix), and \emph{Irrelevant} (different level-1 prefix), with scores $3$, $2$, $1$, and $0$, respectively. Thus, a single representation defines semantic neighborhoods, evaluation instances, and graded relevance, yielding a consistent and reproducible protocol.

\begin{itemize}
\item \textbf{Query--Product Ranking.} Given a natural-language query and a fixed candidate pool containing Exact, Sibling, Related, and Irrelevant items, rank the candidates according to relevance. We report $\mathrm{nDCG@10}$ and $\mathrm{Recall@10}$, both overall and for explicitly constrained attributes.

\item \textbf{Hierarchical Relevance Classification.} Given a query--product pair, predict one of the four relevance classes. We report accuracy and macro-F1.


\item \textbf{Sibling Discrimination.} Given one target item and closely related sibling distractors, identify the target leaf cluster. The task emphasizes fine-grained distinctions between semantically similar products. We report $\mathrm{nDCG@3}$ and $\mathrm{Recall@3}$ for ranked predictions, or binary accuracy when a single item is selected.

\item \textbf{Query Paraphrase Robustness.} We use the query groups generated by our method in Section~\ref{sec:query-gen} as paraphrases: since all queries from a cluster describe the same target intent by construction, we evaluate them against a shared candidate pool. This isolates sensitivity to query wording from candidate-pool variation. We measure robustness using $\mathrm{Consistency@K}$, the mean pairwise Jaccard overlap of top-$K$ retrieved item sets.


\end{itemize}



\section{Evaluation Framework}
\label{sec:evaluation-framework}

FASHION-RQ evaluates retrieval quality at three complementary levels:
(i) task-level performance using the hierarchy-derived relevance labels,
(ii) overall query relevance based on attribute satisfaction, and
(iii) attribute-level diagnostics that identify specific strengths and
failure modes. All metrics are computed from the automatically constructed
queries, query metadata, and SID hierarchy, without requiring
additional human annotation. The four tasks use the common relevance
structure defined in Section~\ref{sec:bench_tasks}.

\noindent \textbf{Task-Level Metrics.}
\label{sec:eval-task-metrics}
We evaluate each task using metrics suited to its corresponding objective as defined in Section~\ref{sec:bench_tasks}.
For \textbf{Query--Product Ranking (T1)}, we report
$\mathrm{nDCG@5/10}$ and $\mathrm{Recall@5/10}$, both overall and for
explicitly constrained attributes. For \textbf{Hierarchical Relevance
Classification (T2)}, we report accuracy and macro-F1, with per-class
precision, recall, and F1 for additional analysis.

For \textbf{Sibling Discrimination (T3)}, we report
$\mathrm{nDCG@1/3}$ and $\mathrm{Recall@1/3}$ to measure fine-grained
target-cluster identification. For \textbf{Query Paraphrase Robustness
(T4)}, we measure retrieval stability using $\mathrm{Consistency@5/10}$,
defined as the mean pairwise Jaccard overlap between the top-$K$ result
sets across paraphrases, and report the mean $\mathrm{nDCG@10}$ across
paraphrases.

\begin{table*}[h]
\centering
\caption{Performance on the four FASHION-RQ tasks. Best results are shown
in bold and second-best results are underlined.
}
\label{tab:main-results}
\resizebox{\textwidth}{!}{
\begin{tabular}{lcc|cc|cc|cc}
\toprule
& \multicolumn{2}{c|}{\textbf{T1: Ranking}}
& \multicolumn{2}{c|}{\textbf{T2: Relevance Classification}}
& \multicolumn{2}{c|}{\textbf{T3: Sibling Discrimination}}
& \multicolumn{2}{c}{\textbf{T4: Paraphrase Robustness}} \\
\textbf{Model}
& \textbf{nDCG@10} & \textbf{Recall@10}
& \textbf{Accuracy} & \textbf{Macro-F1}
& \textbf{nDCG@3} & \textbf{Recall@3}
& \textbf{Consistency@5} & \textbf{Consistency@10} \\
\midrule
BM25~\cite{robertson2009probabilistic}
& 0.7736 & 0.7376
& 0.4023 & 0.3919
& 0.6142 & 0.7723
& 0.5067 & 0.7419 \\

RoBERTa-large~\cite{liu2019roberta}
& 0.7391 & 0.7252
& 0.3749 & 0.3481
& 0.5491 & 0.7260
& 0.4462 & 0.7135 \\

SimCSE-RoBERTa-large~\cite{wu2022smoothedcontrastivelearningunsupervised}
& 0.8655 & 0.8242
& 0.5046 & 0.5047
& 0.6421 & 0.8145
& 0.5914 & 0.8197 \\

Sentence-T5-large~\cite{ni-etal-2022-sentence}
& 0.8235 & 0.7845
& 0.4328 & 0.4227
& 0.6610 & \underline{0.8318}
& 0.5879 & 0.8115 \\

Qwen3-Embedding-4B~\cite{qwen3_embedding_2025}
& \textbf{0.8719} & \textbf{0.8279}
& \textbf{0.5145} & \textbf{0.5146}
& \underline{0.6641} & 0.8232
& \textbf{0.6041} & \textbf{0.8373} \\

Qwen3-Embedding-8B~\cite{qwen3_embedding_2025}
& 0.8686 & 0.8241
& 0.5036 & 0.5037
& 0.6550 & 0.8211
& \underline{0.6022} & \underline{0.8295} \\

E5-Mistral-7B-Instruct~\cite{wang2023improving}
& \underline{0.8699} & \underline{0.8258}
& \underline{0.5134} & \underline{0.5135}
& 0.6462 & 0.8145
& 0.5893 & 0.8233 \\

ColBERTv2~\cite{santhanam2022colbertv2effectiveefficientretrieval}
& 0.8597 & 0.8103
& 0.4827 & 0.4827
& 0.6624 & 0.8242
& 0.5608 & 0.7889 \\

SPLADE++~\cite{formal2021spladev2sparselexical}
& 0.8628 & 0.8141
& 0.4936 & 0.4937
& \textbf{0.6662} & \textbf{0.8349}
& 0.5855 & 0.8089 \\

Contriever-MSMARCO~\cite{izacard2022unsuperviseddenseinformationretrieval}
& 0.8628 & 0.8164
& 0.4796 & 0.4797
& 0.6553 & 0.8277
& 0.5621 & 0.8073 \\
\bottomrule
\end{tabular}
}
\vspace{-4mm}
\end{table*}

\noindent \textbf{Weighted Query Relevance.}
\label{sec:weighted-query-relevance}
The hierarchy-based metrics assess whether retrieved products belong to the
intended semantic neighborhood, but do not quantify satisfaction of the
individual attributes expressed in a query. Moreover, attributes are not
equally important in shopping intent: a user may accept a different brand,
but not a different product type. We therefore introduce a
\emph{weighted query relevance} score that measures query-level intent
satisfaction, with attribute weights derived from a cross-encoder trained to
reflect relevance patterns in real user data.

Let $\mathcal{A}(q)$ denote the attributes explicitly expressed in query $q$,
as specified by its structured representation $\phi(q)$. For each
$a\in\mathcal{A}(q)$, an attribute-match function $m_a(q,p)\in[0,1]$ measures
agreement between the query and product. The comparator depends on attribute
type: exact or synonym matching for categorical attributes, distance-based
matching for ordinal attributes, and semantic similarity for perceptual
attributes. For color, perceptual CIE76~\cite{Robertson1977TheC1} distance is
additionally used when both values are recognized colors. Missing product
attributes receive a neutral score to avoid penalizing incomplete catalog
metadata. Details are provided in the supplementary material.

We aggregate attribute matches using importance weights $w_a$ obtained from
the cross-encoder relevance experiment in Section~\ref{sec:cross-encoder-weights}:
\begin{equation}
R(q,p) = \frac{\sum_{a\in\mathcal{A}(q)}w_a,m_a(q,p)}
{\sum_{a\in\mathcal{A}(q)}w_a}.
\end{equation}
Only explicitly specified attributes contribute, ensuring that $R(q,p)$
measures expressed user intent rather than generic product similarity. The
score lies in $[0,1]$, with higher values indicating stronger overall query
satisfaction.


\noindent\textbf{Attribute-Level Scores.}
\label{sec:attribute-level-scores}
While $R(q,p)$ measures overall query satisfaction, we retain individual
attribute matches to diagnose fine-grained retrieval behavior. For each
attribute $a$, we report the mean match score over queries that explicitly
specify $a$:
$\mathrm{AttrScore}(a)=\frac{1}{|\mathcal{Q}_a|}
\sum_{q\in\mathcal{Q}_a}m_a(q,p_q)$,
where $\mathcal{Q}_a$ denotes the queries specifying $a$. We report scores
for product type, gender, age group, color, material, occasion, brand,
pattern, price, and style, enabling attribute-specific strengths and
failure modes that may be obscured by aggregate metrics.


\section{Experiments}



\noindent \textbf{Cross-encoder derived attribute importance.}
\label{sec:cross-encoder-weights} The attribute weights in $R(q,p)$ are distilled from a cross-encoder trained
on real product-search data, including the ESCI~\cite{reddy2022shopping} filtered to fashion relevant queries and internal production-scale proprietary data. Since its proprietary training
data precludes direct use as an evaluation model, we use the cross-encoder
as a \emph{knowledge source} and estimate attribute importance through
matched twin-pair ablation and regression-based analysis. We normalize the
resulting estimates to obtain non-negative weights $w_a$ that sum to one,
which are used to aggregate per-attribute match scores into the
model-agnostic query-relevance score. The resulting weights are shown in
Table~\ref{tab:query-relevance-attributes}; full estimation procedures,
per-attribute confidence diagnostics, and sample sizes are provided in supplementary.

\noindent \textbf{Baselines.}
\label{sec:baselines} We evaluate FASHION-RQ using ten off-the-\\shelf retrieval baselines spanning
sparse lexical and dense semantic retrieval. The sparse family includes BM25~\cite{robertson2009probabilistic} on
settings. The dense family comprises nine pretrained encoders covering
355M--8B parameter scales, different pooling strategies, and both
non-instruction-tuned and instruction-tuned query encoders. These include
general-purpose sentence encoders as well as recent large embedding models.
All baselines are evaluated \emph{without training or fine-tuning on
FASHION-RQ}; queries and catalog products are encoded using their native
published configurations, and candidates are ranked by lexical matching or
embedding similarity, respectively. This setup evaluates the extent to which
existing retrieval representations generalize to the fine-grained,
hierarchically structured fashion queries in FASHION-RQ.

\begin{table*}[h]
\centering
\caption{Weighted query-relevance and attribute-level matching scores
across retrieval models. Attribute weights $w_a$ are derived from the
cross-encoder and used to compute the overall weighted relevance score, Rel.}
\label{tab:query-relevance-attributes}
\scriptsize
\setlength{\tabcolsep}{7pt}
\renewcommand{\arraystretch}{1.2}
\begin{tabular}{lccccccccccc}
\toprule
\textbf{Model} & \textbf{Rel.} & \textbf{Age} & \textbf{Brand} & \textbf{Color} & \textbf{Gender} & \textbf{Mat.} & \textbf{Occ.} & \textbf{Pattern} & \textbf{Price} & \textbf{Prod.} & \textbf{Style} \\
\midrule
Qwen3-Emb.-8B~\cite{qwen3_embedding_2025} & \textbf{0.9274} & 0.9907 & 0.8560 & \underline{0.9125} & 0.9750 & \textbf{0.5367} & 0.9632 & 0.8521 & 0.6484 & \underline{0.9409} & \underline{0.4291} \\
E5-Mistral-7B~\cite{wang2023improving} & 0.9158 & 0.9900 & \textbf{0.8914} & 0.8782 & 0.9676 & 0.5232 & 0.9618 & 0.8465 & \underline{0.6543} & 0.9395 & 0.4265 \\
Qwen3-Emb.-4B~\cite{qwen3_embedding_2025} & \underline{0.9254} & 0.9895 & 0.8221 & 0.8965 & 0.9721 & 0.5217 & 0.9624 & 0.8327 & 0.6426 & \textbf{0.9438} & 0.4204 \\
Sentence-T5-L~\cite{ni-etal-2022-sentence} & 0.9018 & \underline{0.9918} & 0.7587 & \textbf{0.9159} & \textbf{0.9911} & 0.5310 & 0.9628 & 0.8395 & \textbf{0.6562} & 0.9392 & 0.4229 \\
Sup-SimCSE-RoBERTa-L~\cite{wu2022smoothedcontrastivelearningunsupervised} & 0.8923 & \textbf{0.9922} & 0.6613 & 0.9008 & \underline{0.9873} & 0.5003 & 0.9616 & 0.8296 & 0.6247 & 0.9368 & \textbf{0.4292} \\
BM25~\cite{robertson2009probabilistic} & 0.8753 & 0.9826 & 0.7927 & 0.8451 & 0.8889 & 0.4987 & 0.9585 & 0.7934 & 0.6169 & 0.8929 & 0.4211 \\
RoBERTa-L~\cite{liu2019roberta} & 0.8340 & 0.9671 & 0.6776 & 0.7388 & 0.8793 & 0.3361 & 0.9338 & 0.6373 & 0.5193 & 0.8528 & 0.4222 \\
ColBERTv2~\cite{santhanam2022colbertv2effectiveefficientretrieval} & 0.9198 & 0.9885 & 0.8571 & 0.8615 & 0.9672 & \underline{0.5307} & \underline{0.9638} & \textbf{0.8560} & 0.6228 & 0.9350 & 0.4271 \\
SPLADE++~\cite{formal2021spladev2sparselexical} & \textbf{0.9209} & 0.9877 & \underline{0.8597} & 0.8594 & 0.9712 & \textbf{0.5343} & \textbf{0.9639} & \underline{0.8526} & 0.6325 & 0.9361 & 0.4278 \\
Contriever-MSMARCO~\cite{izacard2022unsuperviseddenseinformationretrieval} & 0.9155 & 0.9850 & 0.8615 & 0.8479 & 0.9560 & 0.5268 & 0.9600 & 0.8332 & 0.6294 & 0.9320 & 0.4247 \\
\midrule
\textbf{$w_a$} & -- & 0.066 & 0.211 & 0.138 & 0.127 & 0.063 & 0.008 & 0.060 & 0.024 & 0.290 & 0.012 \\
\bottomrule
\end{tabular}
\vspace{-4mm}
\end{table*}

\noindent \textbf{Results.}
\label{sec:results}
Table~\ref{tab:main-results} reports the primary results across the four
FASHION-RQ tasks, using two complementary metrics for each task. Overall,
dense and learned sparse retrievers consistently outperform the lexical
BM25 baseline, demonstrating the benefit of semantic representations for
fine-grained fashion retrieval.

For \textbf{T1}, \textit{Qwen3-Embedding-4B} achieves the best performance on
both $\mathrm{nDCG@10}$ (0.8719) and $\mathrm{Recall@10}$ (0.8279),
followed closely by E5-Mistral-7B-Instruct and SimCSE-RoBERTa-large.
The dense and learned sparse models form a relatively tight performance
range, while \textit{RoBERTa-large}, which lacks retrieval-specific
training, performs worst.

For \textbf{T2}, \textit{Qwen3-Embedding-4B} again achieves the best
performance, with 0.5145 accuracy and 0.5146 macro-F1, narrowly followed
by E5-Mistral-7B-Instruct. The close agreement between accuracy and
macro-F1 suggests that the relative ordering is not driven by a single
relevance class. \textit{RoBERTa-large} obtains the lowest accuracy
(0.3749), highlighting the importance of retrieval-oriented representation
learning for hierarchical relevance discrimination.

\textbf{T3} is more challenging, requiring discrimination among products
from closely related sibling clusters. \textit{SPLADE++} achieves the best
$\mathrm{nDCG@3}$ (0.6662) and $\mathrm{Recall@3}$ (0.8349), followed by
\textit{Qwen3-Embedding-4B} on nDCG@3 and \textit{Sentence-T5-large} on
Recall@3. This shows that fine-grained sibling discrimination does not
necessarily improve with model size or dense semantic representations.

For \textbf{T4}, \textit{Qwen3-Embedding-4B} achieves the highest
$\mathrm{Consistency@5}$ (0.6041) and $\mathrm{Consistency@10}$ (0.8373),
indicating the most stable retrieval across paraphrases. Dense and learned
retrievers generally show higher consistency than \textit{BM25}, particularly at
the shallow cutoff, where lexical matching is more sensitive to wording
changes. The weaker ranking but comparatively stable behavior of
\textit{RoBERTa-large} further demonstrates that paraphrase consistency
captures a distinct retrieval property.

Overall, \textit{Qwen3-Embedding-4B} achieves the strongest results on
three of the four tasks, while \textit{SPLADE++} performs best on sibling
discrimination. The complementary results demonstrate that FASHION-RQ
assesses not only overall retrieval quality, but also hierarchical
relevance discrimination and robustness to query variation.

\begin{table}[h]
\centering
\caption{Mean ratings from the user study on a 5-point Likert scale. Higher scores indicate greater perceived naturalness and intent clarity.}
\label{tab:user-study}
\small
\setlength{\tabcolsep}{8pt}
\begin{tabular}{lcc}
\toprule
\textbf{Source} & \textbf{Naturalness} & \textbf{Intent Clarity} \\
\midrule
CHAT  & 3.52 & 2.83 \\
LLM   & 2.84 & 3.76 \\
RQBENCH (Ours) & \textbf{3.62} & \textbf{3.91} \\
\bottomrule
\end{tabular}
\vspace{-2mm}
\end{table}
\noindent \textbf{Query-Relevance Evaluation.} Table~\ref{tab:query-relevance-attributes}
shows that \textit{Qwen3-Embedding-8B} achieves the highest weighted relevance
score (0.9274), followed closely by \textit{SPLADE++} (0.9209) and \textit{ColBERTv2}
(0.9198), demonstrating that both dense and learned sparse retrievers can
capture fine-grained query intent. \textit{BM25} scores lower (0.8753), while
\textit{RoBERTa-Large} ranks last (0.8340). Attribute-level results reveal strong
alignment for \textit{age}, \textit{occasion}, \textit{gender}, and
\textit{product type} across models, whereas \textit{material} and
\textit{style} remain consistently weaker. Notably, the weighted score
distinguishes models with similar aggregate performance through their
attribute-level behavior, demonstrating its ability to expose
fine-grained differences in query satisfaction.

\noindent \textbf{User Study.} \label{sec:user_study} We conducted a user study to assess the \textit{naturalness} and
\textit{intent clarity} of queries from three sources: our \textit{Fashion-RQBENCH} (RQBENCH), real user queries (CHAT), and LLM-generated queries (LLM). 32 participants rated each query on a 5-point Likert scale for both dimensions. Inter-rater reliability was high, with Cronbach's~\cite{Cronbach_1951} $\alpha=0.977$ for naturalness and $\alpha=0.969$ for intent clarity, indicating strong agreement among
raters.

Table~\ref{tab:user-study} summarizes the mean ratings. Naturalness ratings are comparable between RQBENCH and CHAT, with mean scores of 3.62 and 3.52, respectively. In contrast,
intent clarity shows a strong separation: RQBENCH achieves the highest
mean score (3.91), followed by LLM (3.76), while CHAT is substantially
lower (2.83), with all pairwise differences statistically significant.
These results indicate that RQBENCH queries preserve the
naturalness of real user language while providing substantially
clearer and more actionable intent. We provide more details on the study in supplementary. 

\section{Conclusion}

We introduced \textbf{FASHION-RQ}, an automated benchmark construction framework that leverages hierarchical semantic structure to generate realistic queries and graded relevance sets without manual annotation. Using an RQ-VAE with attribute exclusivity supervision, FASHION-RQ produces coherent fine-grained clusters and exploits parent--sibling relations to construct diverse retrieval scenarios. We instantiate this framework as \textit{FASHION-RQBENCH}, a benchmark for evaluating fashion retrieval models through complementary retrieval tasks, weighted query-relevance evaluation, and attribute-level metrics. Together, FASHION-RQ and FASHION-RQBENCH provide a scalable and semantically grounded alternative to manually curated fashion retrieval benchmarks.
FASHION-RQ depends on catalog quality, the underlying multimodal representation, and LLM-generated queries, which may introduce coverage and linguistic biases. The automatically derived relevance structure also approximates rather than replaces human judgments. Future work will explore human validation, broader domains and attributes, richer semantic relations, and adaptive benchmark generation.
\section{Appendix}

\subsubsection{Codebook Size Ablation Study.}
\label{sec:supp_codebook}
We study the effect of RQ-VAE~\cite{lee2022autoregressive} codebook size on the semantic-ID hierarchy by comparing
codebook sizes of $64$, $128$, and $256$ with three quantization levels, while keeping all
other settings fixed. The results characterize the trade-off between \emph{cluster purity}
and \emph{cluster fragmentation}. As the codebook size increases, the available semantic-ID
space grows from $64^3=262{,}144$ to $256^3=16{,}777{,}216$, substantially exceeding the
$44{,}446$ catalog items. Consequently, larger codebooks produce increasingly fragmented
clusters, reducing the number of items sharing the same semantic-ID prefixes.

\begin{table}[h]
\centering
\caption{Effect of codebook size on weighted attribute purity for 64, 128 and 256 size. }
\label{tab:codebook_weighted_purity}
\begin{tabular}{lccc}
\toprule
\textbf{Attribute} & \textbf{64} & \textbf{128} & \textbf{256} \\
\midrule
Master Category & 99.8\% & 50.4\% & 1.0\% \\
Subcategory     & 99.4\% & 50.2\% & 1.0\% \\
Article Type    & 98.3\% & 49.7\% & 1.0\% \\
Gender          & 98.1\% & 49.6\% & 1.0\% \\
Base Color     & 87.9\% & 44.5\% & 1.0\% \\
Usage           & 97.7\% & 49.4\% & 1.0\% \\
\bottomrule
\end{tabular}

\end{table}

Interestingly, weighted global purity and mean per-cluster purity exhibit different trends.
Weighted purity emphasizes the large clusters and therefore decreases sharply with increasing
codebook size. In contrast, mean per-cluster purity can increase because fragmentation creates
many small, trivially pure clusters. Thus, high unweighted cluster purity alone does not imply
a useful semantic hierarchy: excessively large codebooks can produce highly pure but
near-singleton clusters that provide little shared semantic structure.

\begin{table*}[h]
\centering
\caption{Cluster fragmentation as a function of codebook size.}
\label{tab:codebook_density}
\begin{tabular}{lcccccc}
\toprule
& \multicolumn{2}{c}{\textbf{64}} &
\multicolumn{2}{c}{\textbf{128}} &
\multicolumn{2}{c}{\textbf{256}} \\
\textbf{Level} & Clusters & Items/Cl. &
Clusters & Items/Cl. &
Clusters & Items/Cl. \\
\midrule
D1 & 44 & 823.9 & 156 & 285.0 & 267 & 166.5 \\
D2 & 3,378 & 219.2 & 10,378 & 48.28 & 17,378 & 3.56 \\
D3 & 23,956 & 79.5 & 30,045 & 13.48 & 36,122 & 1.23 \\
\bottomrule
\end{tabular}

\end{table*}

The fragmentation trend is evident across all three levels (Table~\ref{tab:codebook_density}).
At the coarsest level, the average cluster size decreases from $824$ items for the
64-codebook to $166.5$ for the 256-codebook. At full depth, the corresponding cluster size
decreases from $1.86$ to $1.23$ items, indicating that many semantic IDs become effectively
item-specific. The 128-codebook configuration provides an intermediate operating point, with
an estimated $285.0$ items per D1 cluster and $1.48$ items per full-depth cluster.

Overall, these results motivate using a moderate codebook size when constructing a semantic-ID
benchmark: the codebook must provide sufficient capacity to separate fine-grained products
while retaining meaningful item sharing within semantic-ID prefixes. In our benchmark
construction, we therefore use the codebook configuration size of 64 which provides the desired balance between semantic purity and
cluster density.

\subsubsection{Query Generation Methodology.}
\label{sec:supp_query_generation}
\paragraph{Signature construction.}
For each semantic-ID prefix, we derive attribute signatures using combinations of
\texttt{masterCategory}, \texttt{subCategory}, \texttt{articleType}, \texttt{brandName},
\texttt{baseColor}, \texttt{gender}, \texttt{usage}, \texttt{price}, and
\texttt{productDisplayName}. We construct two complementary signature sets. 
\emph{Discriminating signatures} identify highly specific cluster characteristics, requiring
at least $10\%$ cluster coverage and $90\%$ precision against the full catalog; the top eight
are retained based on exclusivity, precision, and coverage. These signatures guide query
generation toward attributes that distinguish the target cluster from its siblings.
\emph{Coverage signatures} require at least $5\%$ coverage and $50\%$ precision, with the top
20 retained, and are used to ensure that queries collectively cover the major product types
within the cluster. The primary product-type axis is selected as the most specific attribute
whose dominant value provides the highest cluster purity, with \texttt{masterCategory} as the
fallback.

\paragraph{Query generation.}
For each qualified prefix, the Gemini prompt combines fixed generation instructions with
cluster-specific evidence. The instructions require natural and complete-sentence queries,
coverage of all product types accounting for at least $5\%$ of the cluster plus a catch-all for
minority types, and discrimination from sibling clusters. Queries are grounded in
\texttt{productDisplayName} and description language while avoiding queries based solely on an
attribute value unless that value is shared by all cluster items. The prompt further provides
the major product types and their frequencies, sibling-level attribute contrasts, and a
structured \texttt{DATA CONTEXT} containing the prefix metadata, attribute signatures,
attribute distributions, parent statistics, and sibling differences.

\paragraph{Query Generation Prompt.}
\begin{quote}
\ttfamily\small
You are an expert analyst of hierarchical fashion semantic ID clusters and a specialist in e-commerce search behavior.

Your task is to understand what a given semantic prefix represents by analyzing:
- High-precision attribute signatures (combinations of attributes strongly associated with the prefix)
- Differences from its parent cluster (what makes it more specific)
- Differences from sibling clusters (what makes it unique and non-overlapping)

Based on this, infer the underlying product concept that this prefix encodes.

Then generate natural-language user queries that a real shopper would type into a fashion recommendation system (e.g., Myntra, Amazon, Flipkart). These queries must:

1. Sound natural and human-like (NOT attribute lists like 'color=black AND material=leather').
2. Reflect real shopping intent (style, occasion, product type, etc.).
3. Be phrased properly.
4. Be EXHAUSTIVE: the set of queries must collectively retrieve EVERY item in the prefix group.
   - Generate at least one distinct query for each major product type (>5\% share) listed in the coverage section.
   - Add a broad catch-all query to cover minor types (<5\% share) that are not individually addressed.
5. Be EXCLUSIVE: no query should retrieve items from sibling prefixes or from outside this cluster.
   - For every query, verify it cannot be satisfied by the dominant attributes of each sibling listed in the exclusivity section.
   - Use target-only discriminating attributes as anchors wherever possible.
6. Prefer commonly used fashion terminology over internal attribute names.
7. Ground each query in the dominant phrases from description/productDisplayName to make the intent explicit.
8. Never use a single attribute value unless 100\% of items share it; describe multi-value ranges collectively.

Additional guidelines:
- Use attribute signatures to ground the meaning.
- Use parent deltas to explain what makes this cluster more specific than its parent.
- Use sibling contrasts and the exclusivity constraints below to sharpen boundaries.

Return ONLY a valid JSON object (no markdown fences, no extra text) with these keys:
\{
  "insights": "<concise explanation of what this prefix represents>",
  "user\_queries": ["<query 1>", "<query 2>", ...],   // 6--12 queries, covering all major AND minor product types
  "delta\_vs\_parent": "<key ways this prefix is more specific than its parent>",
  "delta\_vs\_siblings": "<key distinguishing factors vs sibling prefixes>",
  "validation\_checks": ["<rule 1>", ...],   // one rule per major product type confirming coverage; one rule per sibling confirming exclusivity
  "coverage\_map": \{"<product\_type>": "<which query covers it>", ...\}  // explicit mapping from each major product type to the query index that covers it
\}

EXHAUSTIVE COVERAGE REQUIREMENT — this prefix contains \{N\} distinct product types:
  • \{value\}: \{pct\}\% (\{count\} items)  ← MUST be reachable by at least one query
  ...
  • Minority types (<5\% each): \{names\}  ← cover collectively with a broad query

EXCLUSIVITY CONSTRAINTS — your queries must NOT retrieve items from these sibling prefixes:
  Sibling \{prefix\} (\{size\} items)
    Contains: \{top\_categories\}
    Sibling-only attributes (avoid these in target queries): \{attr=val (+Xpp in sibling); ...\}
    Target-only attributes (safe anchors for exclusivity): \{attr=val (+Xpp in target); ...\}
  ...
  For EVERY query you write, verify it would NOT match the dominant attributes of the sibling prefixes above.

DATA CONTEXT:
\{json: target\_prefix.metadata, target\_prefix.blueprint, target\_prefix.discriminating\_signatures,
     target\_prefix.coverage\_signatures, target\_prefix.common\_attributes,
     target\_prefix.full\_attribute\_distribution, parent.record, parent.category\_deltas\_vs\_parent,
     parent.numeric\_deltas\_vs\_parent, sibling\_comparisons\}
\end{quote}

\paragraph{Generation and validation.}
Queries are generated using \texttt{gemini-2.5-flash} with temperature $0.2$ and top-$p=0.8$.
The requested output contains an interpretation of the cluster, $6$--$12$ user queries,
parent/sibling differences, validation checks, and an explicit product-type-to-query
\texttt{coverage\_map}. Responses are parsed using structured JSON extraction with a
free-text fallback to preserve valid queries from partially malformed outputs. The returned
coverage map is subsequently compared against the required product-type set, and uncovered
types are recorded as coverage gaps.

\paragraph{Prefix qualification.}
To ensure that query generation operates on semantically coherent groups, a prefix is processed
only when its \texttt{articleType} purity is at least $0.80$ and its cluster contains at least
two items. Prefixes failing either criterion are excluded from generation.

\subsubsection{Per-task Benchmark Statistics}
\label{sec:supp_benchmark_stats}

Table 1 in main paper reports aggregate statistics for Fashion-RQBENCH;
here we break these down per evaluation task and report the relevance-label
balance, attribute-tag taxonomy, and catalog coverage underlying each task.

\begin{table}[h]
\centering
\caption{Per-task composition of Fashion-RQBENCH. All queries are anchored at
level-3 (leaf) clusters; Task~1 and Task~4 share the same 2,477 anchor queries
(Task~4 paraphrases each one 2--10$\times$).}
\label{tab:benchmark_task_stats}
\resizebox{\linewidth}{!}{
\begin{tabular}{lccccc}
\toprule
\textbf{Task} & \textbf{\# Queries} & \textbf{\# Pairs/Cands} & \textbf{Cands per Query} & \textbf{Leaf Clusters} & \textbf{Query Length} \\
\midrule
T1: Ranking                  & 2,477 & 36,782 & 12--15 (mean 14.8) & 416 & 3--26 words (mean 10.1) \\
T2: Relevance Classification & 2,477 & 19,731 & 7--8 (mean 8.0)     & 416 & 3--26 words (mean 10.1) \\
T3: Sibling Discrimination   & 1,968 & 9,801  & 4--5 (mean 5.0)     & 331 & 3--26 words (mean 10.1) \\
T4: Query Robustness (Paraphrase) & 416 groups & 2,477 queries & 12--15 (mean 14.9) & 416 & 2--10 paraphrases/group (mean 6.0) \\
\bottomrule
\end{tabular}}
\end{table}

\noindent\textbf{Relevance-label balance.} Across the ranking and classification
tasks, candidates are drawn with near-uniform coverage of the four relevance
tiers (Exact / Sibling / Related / Irrelevant), avoiding trivial class-imbalance
shortcuts:

\begin{table}[h]
\centering
\caption{Relevance-label distribution over (query, candidate) pairs.}
\label{tab:relevance_balance}
\resizebox{\linewidth}{!}{
\begin{tabular}{lcccc}
\toprule
\textbf{Task} & \textbf{Exact} & \textbf{Sibling} & \textbf{Related} & \textbf{Irrelevant} \\
\midrule
T1: Ranking (36,782 pairs)         & 20.2\% & 25.9\% & 26.9\% & 26.9\% \\
T2: Classification (19,731 pairs)  & 25.1\% & 24.7\% & 25.1\% & 25.1\% \\
\bottomrule
\end{tabular}}
\end{table}

\noindent\textbf{Attribute-tag taxonomy.} Each ranking query is annotated with
0--8 attribute tags (mean 3.8) drawn from a 10-way taxonomy, with the following
coverage over the 2,477 queries: \textit{product\_type} (2,475), \textit{gender}
(1,892), \textit{occasion} (1,349), \textit{brand} (945), \textit{style\_tags}
(896), \textit{color} (845), \textit{age\_group} (627), \textit{pattern} (156),
\textit{material} (153), \textit{price\_hint} (80).

\noindent\textbf{Sibling discrimination.} Each of the 1,968 queries pairs
exactly one correct item against 3--4 distractor clusters (mean 3.89) sampled
from neighboring leaf clusters, yielding 4--5 candidates per query — a
harder, lower-cardinality setting than Task~1/2 designed to isolate
fine-grained (sibling-level) confusability.

\noindent\textbf{Catalog coverage.} The benchmark draws from the full
44,446-item catalog with broad metadata diversity: sampled items span
\textit{masterCategory} \{Apparel: 7,622, Accessories: 6,760, Footwear: 3,699,
Personal Care: 1,570, Free Items: 73, Sporting Goods: 7\} (Task~2 pairs) and
\textit{gender} \{Men: 8,996, Women: 8,913, Unisex: 1,279, Boys: 284,
Girls: 259\}, with price ranging \$90--\$17,995 (median \$1,199, mean \$1,744).
Across all four tasks, the benchmark touches 8,841 unique items (19.9\% of the
catalog) as either queries' anchor items or candidates/distractors.

\subsubsection{Attribute-Level Relevance Scoring}
\label{sec:supp_attribute_scoring}

This section details the comparator $m_a(q,p)$ used for each attribute
$a\in\mathcal{A}(q)$ in weighted query relevance section of the main paper. All soft-text
comparators use cosine similarity between Marqo-FashionCLIP~\cite{marqo2024fashionsiglip} text embeddings
of the compared strings, denoted $\mathrm{sim}(\cdot,\cdot)$. When a product
lacks a structured value for an expressed query attribute, the comparator
returns a neutral score of $0.5$ rather than $0$, so that missing catalog
metadata is not conflated with an actual mismatch.

\paragraph{Product type.} A hierarchical comparison against the product's
full category path $\mathrm{masterCategory} \!>\! \mathrm{subCategory} \!>\!
\mathrm{articleType}$. If the query's product-type string occurs as a
substring anywhere in the lower-cased path, the score is floored at $0.85$
(refined upward by $\mathrm{sim}$ against the leaf category when available).
Otherwise, $m_a(q,p) = \max\big(\max_{\ell \in \text{levels}}
\mathrm{sim}(q_a, \ell),\ \mathrm{sim}(q_a, \text{full path})\big)$, i.e.\ the
best similarity across the three individual hierarchy levels and the
concatenated path.

\paragraph{Gender.} Categorical, after synonym normalization (e.g.\ "men",
"man", "male", "boys" $\to$ \textit{male}). Exact match after normalization
scores $1.0$; a query or product value of \textit{unisex} against the other
gender scores $0.7$; any other mismatch scores $0.1$.

\paragraph{Age group.} Ordinal, over the ordered scale
$\{\text{baby}, \text{toddler}, \text{kids}, \text{adult}\}$ (Myntra's
\textit{ageGroup} field, e.g.\ "Adults-Men", is reduced to this scale by
keeping only the age segment). Exact match scores $1.0$; otherwise the score
decays linearly with ordinal distance, $\max(0, 1 - 0.35\,|\,\mathrm{idx}(q_a)
- \mathrm{idx}(p_a)\,|)$; a value outside the ordered scale scores $0.3$.

\paragraph{Color.} Perceptual CIE76 $\Delta E$ distance in CIELAB~\cite{Robertson1977TheC1} space between a query color and a product color, computed via a curated
dictionary of $\sim$90 named colors mapped to representative sRGB values.
Identical color strings score $1.0$ directly. When both values resolve to a
known color, $m_a(q,p) = \max(0,\ 1 - \Delta E / 120)$, taking the maximum
over all query$\times$product color pairs. When either color is not in the
dictionary, the comparator falls back to $\mathrm{sim}(q_a, p_a)$. If the
product has no structured \textit{baseColor}, the query color list is
instead compared against the product description text.

\paragraph{Material, occasion, pattern, style.} These four attributes share
one comparator shape: the product's value list is parsed from the
semi-structured \textit{articleAttributes} field (fabric/pattern/fit/occasion
keys) and, when that yields nothing, backfilled by matching a curated
domain vocabulary (e.g.\ pattern terms like "floral print", "polka dot";
style/fit terms like "slim fit", "off shoulder"; occasion terms like
"business casual", "loungewear") against the product title and description.
Given the resulting product value list, $m_a(q,p) = \max_{v\in q_a,\ u\in
p_a} \mathrm{sim}(v, u)$ over all query$\times$product value pairs. If the
product has no values for the attribute at all, $q_a$ is instead compared
against the raw description (and, for occasion/pattern/style, the full
category path) text as a single string.

\paragraph{Brand.} Case-insensitive exact-or-substring match against the
catalog's brand field: $m_a(q,p) = 1.0$ if either string contains the other,
else $0.0$.

\paragraph{Price hint.} Both query and product price are bucketed into three
tiers (\textit{budget} / \textit{mid-range} / \textit{premium}) using
INR thresholds (\textrm{<}\,700, 700--2000,
\textrm{>}\, 2000, from the catalog's list price). Matching tiers
score $1.0$; any mismatch scores $0.3$ (rather than $0$, since adjacent price
tiers are a much smaller miss than an unrelated product).

\paragraph{Optional visual grounding.} When enabled, product\_type, color,
pattern, material, occasion, and style additionally probe a FashionCLIP
image embedding of the product photo with a short attribute-specific text
prompt (the attribute value plus the query's product type as context, e.g.\
"striped shirt"), and the attribute's final score is
$\max(m_a^{\text{text}}(q,p),\ m_a^{\text{image}}(q,p))$ — letting either
signal recover from a missing or noisy metadata field.

\paragraph{Aggregation.} For a given query, only the subset of attributes
$\mathcal{A}(q)$ actually expressed in that query's structured representation
contribute to its relevance score; the corresponding cross-encoder weights
$w_a$ are renormalized to sum to $1$ over $\mathcal{A}(q)$ before the
weighted sum is taken. A query with a multi-intent structured representation
(e.g.\ a gift-set query implying several distinct product asks) is scored
against every intent variant independently, and the best-matching variant's
score is kept — a retrieved product is credited if it satisfies at least one
of the query's intents well.

\subsubsection{Cross-Encoder-Based Attribute Importance Estimation}
\label{sec:supp_cross_encoder}

We estimate attribute-level importance using a cross-encoder trained on real fashion search
interactions. The cross-encoder serves as an independent relevance oracle, allowing us to
derive attribute weights without relying on the retrieval systems or evaluation metrics
being compared in Fashion-RQBENCH. We use two complementary estimation procedures: a
controlled matched-pair ablation and an independent regression-based cross-check.

\paragraph{Cross-encoder oracle.}

 We train a cross-encoder using the ModernBERT-large~\cite{warner-etal-2025-smarter} architecture on a combination of proprietary internal relevance dataset (prop-rel) and a fashion-filtered subset of ESCI~\cite{reddy2022shopping}. The model predicts binary query--product relevance, and
 we use its raw output logit as the relevance
 score. As ESCI's US-locale subset spans fashion and non-fashion domains, we filter it to (query, product) pairs whose title or description contains fashion-specific keywords, treating products labeled \emph{exact} for a given query as relevant and those labeled \emph{substitute}, \emph{complement}, or \emph{irrelevant} as irrelevant; a query may be paired with multiple products, yielding multiple query-product pairs. Combining this ESCI-filtered subset with prop-rel gives 539{,}849 training samples over 44{,}670 unique queries and 58{,}792 validation samples over 4{,}985 unique queries. Of the training queries, 14{,}467 (32.4\%) come from the ESCI-filtered set and 30{,}203 (67.6\%) from prop-rel; the validation split has an analogous 1{,}611 (32.3\%) / 3{,}374 (67.7\%) split. At the (query, product) sample level, ESCI accounts for 251{,}100 (46.5\%) of training data and 27{,}876 (47.4\%) of validation data, with the remainder — 288{,}749 (53.5\%) train, 30{,}916 (52.6\%) val — from prop-rel. 
 
 We initialize the model from a checkpoint ~\cite{moderncestsb2025} pretrained on sentence similarity, and represent product text as \text{"name \textbar brand \textbar gender \textbar color \textbar materials \textbar description"}. The model is trained with binary cross-entropy loss, mean pooling, label smoothing ($\epsilon=0.1$), and batch size 512. On our internal test set of 2{,}353 unique queries and 12,619 query-product pairs (positive rate 0.632), with threshold=0.5, the cross-encoder achieves 0.818 ROC-AUC and a Brier score of 0.162, versus 0.233 for a baseline predicting the marginal positive rate for every pair.
 
 We validate the oracle on a manually constructed sanity set of 100 query--product examples covering multiple attributes. The model
achieves a relevant--irrelevant probability margin of $0.616$ and pairwise ranking accuracy
of $97.0\%$, exceeding our validation criteria of $0.15$ and $75\%$, respectively.

\paragraph{Attribute annotations.}
Item-level attributes are obtained independently from the public Myntra fashion catalog~\cite{param_aggarwal_2019}.
Product type, gender, age group, color, and usage/occasion are obtained from structured
metadata, while material and pattern are extracted from the attribute text. Price is
discretized into budget, mid-range, and premium categories. Query-side attributes are obtained
from naturally occurring enriched queries. Query and item attributes are canonicalized using
the same normalization procedure before computing attribute matches.

\paragraph{Matched-pair ablation.}
For each attribute $a$, we construct matched ``twin'' item pairs that agree on relevant
control attributes but differ in $a$. Given a query specifying value $v_1$ of attribute $a$,
the two items are scored by the cross-encoder and their logit difference is computed as
\begin{equation}
\Delta_i =
s_{\mathrm{CE}}(q_i,x_1)-s_{\mathrm{CE}}(q_i,x_2).
\end{equation}
We additionally score each twin pair against randomly sampled queries to obtain a null
distribution. For each attribute, we report the mean logit difference, a twin-pair-clustered
95\% bootstrap confidence interval using 2,000 resamples, sign consistency, and a
distinguishability-from-null diagnostic. Confidence tiers reflect both sample size and
whether the confidence interval excludes zero. Results are shown in
Table~\ref{tab:ce-method-a}.

\begin{table*}[t]
\centering
\caption{Controlled matched-twin-pair ablation using the cross-encoder relevance oracle.
$N$ denotes the number of query--twin-pair observations. $\overline{\Delta}$ is the mean
cross-encoder logit difference, and the confidence interval is obtained by resampling
twin pairs with replacement. Sign consistency denotes the fraction of observations having
the same sign as $\overline{\Delta}$.}
\label{tab:ce-method-a}
\begin{tabular}{lccccc}
\toprule
\textbf{Attribute} &
\textbf{$N$} &
\textbf{$\overline{\Delta}$} &
\textbf{95\% CI} &
\textbf{Sign Consistency} \\
\midrule
Product Type & 561 & 0.579 & [0.440, 0.713] & 0.78 \\
Gender       & 730 & 0.730 & [0.625, 0.838] & 0.80  \\
Age Group    & 129 & 0.737 & [0.537, 0.951] & 0.83  \\
Color        & 786 & 0.926 & [0.828, 1.025] & 0.86  \\
Material     & 259 & 0.589 & [0.384, 0.796] & 0.75  \\
Occasion     & 952 & 0.072 & [$-$0.010, 0.153] & 0.53 \\
Pattern      & 282 & 0.581 & [0.342, 0.830] & 0.71 \\
Style Tags   & 6   & 0.002 & [$-$0.274, 0.513] & 0.50 \\
Brand        & 633 & 1.409 & [1.282, 1.537] & 0.90  \\
Price Hint   & 172 & 0.242 & [0.046, 0.433] & 0.60  \\
\bottomrule
\end{tabular}
\end{table*}

\paragraph{Regression cross-check.}
As an independent validation, we construct query--product pairs from real queries and encode
each attribute as $f_a\in\{1,0,0.5\}$ for match, mismatch, and unspecified, respectively.
After standardizing the ten attribute features, we fit a Ridge regression~\cite{vanwieringen2026lecturenotesridgeregression} to the
cross-encoder logits using $N=6{,}205$ pairs. The regression achieves $R^2=0.490$.
Variance inflation factors (VIFs)~\cite{marquardt1970generalized} between $1.00$ and $1.10$ indicate negligible
multicollinearity among the attribute features. The regression coefficients, VIFs, and
resulting blended weights are reported in Table~\ref{tab:ce-method-b}.

\begin{table}[t]
\centering
\caption{Regression-based attribute importance and final blended weights. Ridge coefficients
are estimated from cross-encoder logits. VIF measures multicollinearity among attribute
features.}
\label{tab:ce-method-b}
\begin{tabular}{lccc}
\toprule
\textbf{Attribute} & \textbf{Ridge Coef.} & \textbf{VIF} & \textbf{Final $w_a$} \\
\midrule
Product Type & 0.467 & 1.10 & 0.290 \\
Gender       & 0.125 & 1.06 & 0.127 \\
Age Group    & 0.006 & 1.00 & 0.066 \\
Color        & 0.114 & 1.01 & 0.138 \\
Material     & 0.026 & 1.02 & 0.063 \\
Occasion     & 0.004 & 1.02 & 0.008 \\
Pattern      & 0.021 & 1.00 & 0.060 \\
Style Tags   & 0.024 & 1.02 & 0.012 \\
Brand        & 0.177 & 1.02 & 0.211 \\
Price Hint   & 0.006 & 1.00 & 0.024 \\
\bottomrule
\end{tabular}
\end{table}

\paragraph{Weight construction.}
For each method, negative estimates are clipped to zero and the resulting values are
normalized to sum to one. We compute the Spearman rank correlation between the two sets of
raw estimates as an agreement measure. The final attribute weight $w_a$ is obtained by
averaging the two normalized weight vectors and renormalizing the result to sum to one.
These weights are subsequently used by the attribute-aware relevance metric in
Section 3 of main paper.

\subsubsection{Baseline Models and Implementation Details}
\label{sec:supp_baselines}

\subsection{Baseline Model Configurations}
\label{sec:supp_model_configs}

We evaluate 12 retrieval baselines spanning four architecture families---lexical,
learned-sparse, multi-vector late-interaction, and dense single-vector models---following
the taxonomy of~\citet{kamalloo2024resources}. All models are evaluated as
\emph{frozen, off-the-shelf} checkpoints without task-specific fine-tuning. The exact
configurations used for each baseline are provided in Tables~\ref{tab:models_lexical}--\ref{tab:models_llm}.

\begin{table}[h]
\centering
\caption{Lexical, learned-sparse, and multi-vector retrieval baselines.}
\label{tab:models_lexical}
\resizebox{\linewidth}{!}{%
\begin{tabular}{p{0.28\linewidth}p{0.19\linewidth}p{0.48\linewidth}}
\toprule
\textbf{Model} & \textbf{Scoring} & \textbf{Key Hyperparameters} \\
\midrule

BM25~\cite{robertson2009probabilistic}
& Sparse lexical
& $k_1=1.2$, $b=0.75$; English stopwords; Porter stemmer \\

SPLADE++ EnsembleDistil~\cite{formal2021spladev2sparselexical}
& Learned sparse
& \texttt{max\_seq\_len}=512; batch size=64;
$\max(\log(1+\mathrm{ReLU}(\cdot)))$ over MLM logits;
vocabulary size=30{,}522 \\

ColBERTv2~\cite{santhanam2022colbertv2effectiveefficientretrieval}
& Multi-vector MaxSim
& 128 dimensions/token; \texttt{query\_maxlen}=32;
\texttt{doc\_maxlen}=180; FP16 with FP32 MaxSim accumulation;
batch size=128 \\

\bottomrule
\end{tabular}%
}
\end{table}

\begin{table}[h]
\centering
\caption{Dense single-vector retrieval baselines with non-LLM backbones.}
\label{tab:models_dense}
\resizebox{\linewidth}{!}{%
\begin{tabular}{llcccl}
\toprule
\textbf{Model}  & \textbf{Dim.} & \textbf{Pooling} &
\textbf{Max Seq. Len}  \\
\midrule

Contriever-MSMARCO~\cite{izacard2022unsuperviseddenseinformationretrieval}
& 768
& Mean
& 512
 \\

RoBERTa-large~\cite{liu2019roberta}
& 1024
& Mean(first + last layer)
& 512
 \\

SimCSE-sup (RoBERTa-large)~\cite{wu2022smoothedcontrastivelearningunsupervised}
& 1024
& \texttt{pooler\_output} (CLS)
& 512
 \\

Sentence-T5-large~\cite{ni-etal-2022-sentence}
& 768$^{\dagger}$
& Mean + Dense($1024\!\rightarrow\!768$, tanh)
& 256
 \\

\bottomrule
\end{tabular}%
}
\end{table}

\noindent
\textsuperscript{$\dagger$}The base T5-large hidden size is 1024; the model's built-in
projection head produces 768-dimensional embeddings.

\begin{table}[h]
\centering
\caption{LLM-backbone embedding baselines. All models use last-token pooling,
\texttt{bfloat16}, and $\ell_2$ normalization. Queries are prefixed with
\emph{``Given a fashion product search query, retrieve relevant product listings that match
the query''}, while documents are unprefixed.}
\label{tab:models_llm}
\resizebox{\linewidth}{!}{
\begin{tabular}{llccccc}
\toprule
\textbf{Model} & \textbf{Params.} & \textbf{Dim.} &
\textbf{Max Seq. Len} & \textbf{EOS Policy}  \\
\midrule
Qwen3-Embedding-0.6B~\cite{qwen3_embedding_2025}  &
0.6B & 1024 & 8192 & auto$^\ddagger$ \\
Qwen3-Embedding-4B~\cite{qwen3_embedding_2025}  &
4B & 2560 & 8192 & auto$^\ddagger$ \\
Qwen3-Embedding-8B~\cite{qwen3_embedding_2025}  &
8B & 4096 & 8192 & auto$^\ddagger$  \\
E5-Mistral-7B-Instruct~\cite{wang2023improving}  &
7B & 4096 & 4096 & manual$^\S$  \\
SFR-Embedding-Mistral~\cite{SFRAIResearch2024}  &
7B & 4096 & 4096 & auto  \\
\bottomrule
\end{tabular}}
\end{table}

\noindent
\textsuperscript{$\ddagger$}\texttt{auto} EOS handling was empirically verified against the
tokenizer for Qwen3-Embedding-0.6B and the same policy was applied to the 4B and 8B
checkpoints without independent re-verification. \textsuperscript{$\S$}For
E5-Mistral-7B-Instruct, EOS is appended once manually after truncation.
.

\section{User Study Additional Details}
\label{sec:user-study-supp}

\paragraph{Query sources.}
We evaluated three blinded query sources, with 20 queries per source ($N=60$):
\textbf{RQBENCH} sampled our generated queries; \textbf{CHAT}, real user shopping queries filtered by majority-vote intent labels ($\geq 2/3$ LLM judges), scope, length, deduplication, and manual NSFW/PII screening; and \textbf{LLM}, unconditioned, persona-free shopping-query generations without dataset conditioning or few-shot examples. For RQBENCH, 20 queries were randomly sampled  (seed 42). CHAT and LLM were similarly sampled from curated pools of 50 using seed 42. All 60 queries were then randomly interleaved (seed 2026), assigned hidden IDs, and presented without source labels.

\paragraph{Rating protocol.}
Each query was rated independently on two 5-point Likert scales:
\textbf{naturalness} (1 = very unnatural/scripted, 5 = very natural) and
\textbf{intent clarity} (1 = very vague, 5 = very clear/specific). Raters were explicitly instructed to treat the dimensions independently.

\paragraph{Participants and reliability.}
Thirty-two raters evaluated all 60 queries on both dimensions, yielding valid ratings after removing unparseable responses. Cronbach's $\alpha$ was 0.977 for naturalness and 0.969 for intent clarity.

\paragraph{Statistical analysis.}
For each dimension, we performed a Kruskal--Wallis omnibus test followed by pairwise Mann--Whitney $U$ tests for all three source pairs. Pairwise $p$-values were Holm-corrected, with $\alpha=0.05$.

\paragraph{Qualitative examples.}
Representative naturalness scores (lowest/highest) were: RQBENCH, \textit{`Casual black or brown sunglasses.''} (3.10) / \textit{`Celine Dion brown sunglasses.''} (4.03); CHAT, \textit{`Put me in a dusty pink lilac saree with like 50\% sheer flowy fabric and a light golden blouse with straps sweetheart neckline.''} (2.52) / \textit{`Silk gown.''} (4.23); and LLM, \textit{`Can you show me some elegant evening gowns for a black-tie event?''} (2.77) / \textit{`Show me some stylish handbags for everyday use.''} (4.33).

 Representative intent-clarity scores (lowest/highest) were: RQBENCH, \textit{I want to buy a perfume gift set.''} (3.37) / \textit{Show me Celine Dion brown sunglasses.''} (4.31); CHAT, \textit{Dark green pieces.''} (1.66) / \textit{Put me in a dusty pink lilac saree with like 50\% sheer flowy fabric and a light golden blouse with straps sweetheart neckline.''} (4.00); and LLM, \textit{I need a dress that transitions from day to night.''} (2.33) / \textit{I want a raincoat that's both stylish and waterproof.''} (4.29).

{
    \small
    \bibliographystyle{ieeenat_fullname}
    \bibliography{main}

@String(CVPR= {IEEE Conf. Comput. Vis. Pattern Recog.})

@String(ICCV= {Int. Conf. Comput. Vis.})

@String(ECCV= {Eur. Conf. Comput. Vis.})

@String(NIPS= {Adv. Neural Inform. Process. Syst.})

@String(AAAI = {AAAI})

@String(CVPR  = {CVPR})

@String(ICCV  = {ICCV})

@String(ECCV  = {ECCV})

@String(NIPS  = {NeurIPS})

@inproceedings{liu2016deepfashion,
 author = {Liu, Ziwei and Luo, Ping and Qiu, Shi and Wang, Xiaogang and Tang, Xiaoou},
 title = {DeepFashion: Powering Robust Clothes Recognition and Retrieval with Rich Annotations},
 booktitle = {Proceedings of IEEE Conference on Computer Vision and Pattern Recognition (CVPR)},
 month = {June},
 year = {2016} 
}

@inproceedings{ge2019deepfashion2,
  title={Deepfashion2: A versatile benchmark for detection, pose estimation, segmentation and re-identification of clothing images},
  author={Ge, Yuying and Zhang, Ruimao and Wang, Xiaogang and Tang, Xiaoou and Luo, Ping},
  booktitle={2019 IEEE/CVF Conference on Computer Vision and Pattern Recognition (CVPR)},
  pages={5332--5340},
  year={2019},
  organization={IEEE}
}

@inproceedings{han2017fashion200k,
  title={Automatic spatially-aware fashion concept discovery},
  author={Han, Xintong and Wu, Zuxuan and Huang, Phoenix X and Zhang, Xiao and Zhu, Menglong and Li, Yuan and Zhao, Yang and Davis, Larry S},
  booktitle={2017 IEEE international conference on computer vision (ICCV)},
  pages={1472--1480},
  year={2017},
  organization={IEEE}
}

@inproceedings{yang2020fashioncaptioning,
author="Yang, Xuewen
and Zhang, Heming
and Jin, Di
and Liu, Yingru
and Wu, Chi-Hao
and Tan, Jianchao
and Xie, Dongliang
and Wang, Jue
and Wang, Xin",
editor="Vedaldi, Andrea
and Bischof, Horst
and Brox, Thomas
and Frahm, Jan-Michael",
title="Fashion Captioning: Towards Generating Accurate Descriptions with Semantic Rewards",
booktitle="Computer Vision -- ECCV 2020",
year="2020",
publisher="Springer International Publishing",
address="Cham",
pages="1--17",
isbn="978-3-030-58601-0"
}

@inproceedings{ak2018shopping100k,
  author={Ak, Kenan E. and Lim, Joo Hwee and Tham, Jo Yew and Kassim, Ashraf A.},
  booktitle={2018 IEEE Winter Conference on Applications of Computer Vision (WACV)}, 
  title={Efficient Multi-attribute Similarity Learning Towards Attribute-Based Fashion Search}, 
  year={2018},
  volume={},
  number={},
  pages={1671-1679},
  doi={10.1109/WACV.2018.00186}}

@inproceedings{wu2021fashioniq,
  title={Fashion iq: A new dataset towards retrieving images by natural language feedback},
  author={Wu, Hui and Gao, Yupeng and Guo, Xiaoxiao and Al-Halah, Ziad and Rennie, Steven and Grauman, Kristen and Feris, Rogerio},
  booktitle={2021 IEEE/CVF Conference on Computer Vision and Pattern Recognition (CVPR)},
  pages={11302--11312},
  year={2021},
  organization={IEEE}
  }

@inproceedings{liu2021cirr,
    title={Image retrieval on real-life images with pre-trained vision-and-language models},
  author={Liu, Zheyuan and Rodriguez-Opazo, Cristian and Teney, Damien and Gould, Stephen},
  booktitle={2021 IEEE/CVF International Conference on Computer Vision (ICCV)},
  pages={2105--2114},
  year={2021},
  organization={IEEE}
}

@inproceedings{baldrati2023searle,
title={Zero-shot composed image retrieval with textual inversion},
  author={Baldrati, Alberto and Agnolucci, Lorenzo and Bertini, Marco and Del Bimbo, Alberto},
  booktitle={2023 IEEE/CVF International Conference on Computer Vision (ICCV)},
  pages={15292--15301},
  year={2023},
  organization={IEEE}
}

@inproceedings{levy2024lasco,
  title={Data roaming and quality assessment for composed image retrieval},
  author={Levy, Matan and Ben-Ari, Rami and Darshan, Nir and Lischinski, Dani},
  booktitle={Proceedings of the AAAI conference on artificial intelligence},
  volume={38},
  number={4},
  pages={2991--2999},
  year={2024}
}

@article{chia2022fashionclip,
  title={Contrastive language and vision learning of general fashion concepts},
  author={Chia, Patrick John and Attanasio, Giuseppe and Bianchi, Federico and Terragni, Silvia and Magalhaes, Ana Rita and Goncalves, Diogo and Greco, Ciro and Tagliabue, Jacopo},
  journal={Scientific Reports},
  volume={12},
  number={1},
  pages={18958},
  year={2022},
  publisher={Nature Publishing Group UK London}
}

@software{marqo2024fashionsiglip,
author = {Jung, Myong Chol and Clark, Jesse},
month = aug,
title = {{Marqo-FashionCLIP and Marqo-FashionSigLIP}},
url = {https://github.com/marqo-ai/marqo-FashionCLIP},
version = {1.0.0},
year = {2024}
}

@inproceedings{zhu2024gcl,
  title={Generalized contrastive learning for multi-modal retrieval and ranking},
  author={Zhu, Tianyu and Jung, Myong Chol and Clark, Jesse},
  booktitle={Companion Proceedings of the ACM on Web Conference 2025},
  pages={661--670},
  year={2025}
}

@article{reddy2022esci,
  title={Shopping queries dataset: A large-scale ESCI benchmark for improving product search},
  author={Reddy, Chandan K and M{\`a}rquez, Llu{\'\i}s and Valero, Fran and Rao, Nikhil and Zaragoza, Hugo and Bandyopadhyay, Sambaran and Biswas, Arnab and Xing, Anlu and Subbian, Karthik},
  journal={arXiv preprint arXiv:2206.06588},
  year={2022}
}

@article{tay2022dsi,
  title={Transformer memory as a differentiable search index},
  author={Tay, Yi and Tran, Vinh and Dehghani, Mostafa and Ni, Jianmo and Bahri, Dara and Mehta, Harsh and Qin, Zhen and Hui, Kai and Zhao, Zhe and Gupta, Jai and others},
  journal={Advances in neural information processing systems},
  volume={35},
  pages={21831--21843},
  year={2022}
}

@article{wang2022nci,
  title={A neural corpus indexer for document retrieval},
  author={Wang, Yujing and Hou, Yingyan and Wang, Haonan and Miao, Ziming and Wu, Shibin and Chen, Qi and Xia, Yuqing and Chi, Chengmin and Zhao, Guoshuai and Liu, Zheng and others},
  journal={Advances in Neural Information Processing Systems},
  volume={35},
  pages={25600--25614},
  year={2022}
}

@article{sun2023genret,
  title={Learning to tokenize for generative retrieval},
  author={Sun, Weiwei and Yan, Lingyong and Chen, Zheng and Wang, Shuaiqiang and Zhu, Haichao and Ren, Pengjie and Chen, Zhumin and Yin, Dawei and Rijke, Maarten and Ren, Zhaochun},
  journal={Advances in Neural Information Processing Systems},
  volume={36},
  pages={46345--46361},
  year={2023}
}

@inproceedings{zeng2024ripor,
  title={Scalable and effective generative information retrieval},
  author={Zeng, Hansi and Luo, Chen and Jin, Bowen and Sarwar, Sheikh Muhammad and Wei, Tianxin and Zamani, Hamed},
  booktitle={Proceedings of the ACM Web Conference 2024},
  pages={1441--1452},
  year={2024}
}

@article{rajput2023tiger,
  title={Recommender systems with generative retrieval},
  author={Rajput, Shashank and Mehta, Nikhil and Singh, Anima and Hulikal Keshavan, Raghunandan and Vu, Trung and Heldt, Lukasz and Hong, Lichan and Tay, Yi and Tran, Vinh and Samost, Jonah and others},
  journal={Advances in Neural Information Processing Systems},
  volume={36},
  pages={10299--10315},
  year={2023}
}

@inproceedings{singh2023semanticids,
  title={Better generalization with semantic ids: A case study in ranking for recommendations},
  author={Singh, Anima and Vu, Trung and Mehta, Nikhil and Keshavan, Raghunandan and Sathiamoorthy, Maheswaran and Zheng, Yilin and Hong, Lichan and Heldt, Lukasz and Wei, Li and Tandon, Devansh and others},
  booktitle={Proceedings of the 18th ACM Conference on Recommender Systems},
  pages={1039--1044},
  year={2024}
}

@inproceedings{lee2022autoregressive,
  title={Autoregressive image generation using residual quantization},
  author={Lee, Doyup and Kim, Chiheon and Kim, Saehoon and Cho, Minsu and Han, Wook-Shin},
  booktitle={2022 IEEE/CVF conference on computer vision and pattern recognition (CVPR)},
  pages={11513--11522},
  year={2022},
  organization={IEEE}
}

@inproceedings{oord2017vqvae,
  title     = {Neural Discrete Representation Learning},
  author    = {van den Oord, Aaron and Vinyals, Oriol and Kavukcuoglu, Koray},
  booktitle = NIPS,
  year      = {2017}
}

@article{nogueira2019doc2query,
  title={Document expansion by query prediction},
  author={Nogueira, Rodrigo and Yang, Wei and Lin, Jimmy and Cho, Kyunghyun},
  journal={arXiv preprint arXiv:1904.08375},
  year={2019}
}

@article{bonifacio2022inpars,
  title={Inpars: Data augmentation for information retrieval using large language models},
  author={Bonifacio, Luiz and Abonizio, Hugo and Fadaee, Marzieh and Nogueira, Rodrigo},
  journal={arXiv preprint arXiv:2202.05144},
  year={2022}
}

@article{dai2023promptagator,
  title={Promptagator: Few-shot Dense Retrieval From 8 Examples},
  author={Zhuyun Dai and Vincent Zhao and Ji Ma and Yi Luan and Jianmo Ni and Jing Lu and Anton Bakalov and Kelvin Guu and Keith B. Hall and Ming-Wei Chang},
  journal={ArXiv},
  year={2022},
  volume={abs/2209.11755},
  url={https://api.semanticscholar.org/CorpusID:252519173}
}

@article{thakur2021beir,
    title={Beir: A heterogenous benchmark for zero-shot evaluation of information retrieval models},
  author={Thakur, Nandan and Reimers, Nils and R{\"u}ckl{\'e}, Andreas and Srivastava, Abhishek and Gurevych, Iryna},
  journal={arXiv preprint arXiv:2104.08663},
  year={2021}
}

@inproceedings{zhang2024excluir,
  title={Excluir: Exclusionary neural information retrieval},
  author={Zhang, Wenhao and Zhang, Mengqi and Wu, Shiguang and Pei, Jiahuan and Ren, Zhaochun and De Rijke, Maarten and Chen, Zhumin and Ren, Pengjie},
  booktitle={Proceedings of the AAAI conference on artificial intelligence},
  volume={39},
  number={12},
  pages={13295--13303},
  year={2025}
}

@article{jarvelin2002ndcg,
author = {J{\"a}rvelin, Kalervo and Kek{\"a}l{\"a}inen, Jaana},
title = {Cumulated gain-based evaluation of IR techniques},
year = {2002},
issue_date = {October 2002},
publisher = {Association for Computing Machinery},
address = {New York, NY, USA},
volume = {20},
number = {4},
issn = {1046-8188},
url = {https://doi.org/10.1145/582415.582418},
doi = {10.1145/582415.582418},
journal = {ACM Trans. Inf. Syst.},
month = oct,
pages = {422–446},
numpages = {25}
}

@article{upadhyay2024umbrela,
  title={Umbrela: Umbrela is the (open-source reproduction of the) bing relevance assessor},
  author={Upadhyay, Shivani and Pradeep, Ronak and Thakur, Nandan and Craswell, Nick and Lin, Jimmy},
  journal={arXiv preprint arXiv:2406.06519},
  year={2024}
}

@inproceedings{ma2020asen,
  title={Fine-grained fashion similarity learning by attribute-specific embedding network},
  author={Ma, Zhe and Dong, Jianfeng and Long, Zhongzi and Zhang, Yao and He, Yuan and Xue, Hui and Ji, Shouling},
  booktitle={Proceedings of the AAAI Conference on artificial intelligence},
  volume={34},
  number={07},
  pages={11741--11748},
  year={2020}
}

@inproceedings{vaze2023genecis,
  title={Genecis: A benchmark for general conditional image similarity},
  author={Vaze, Sagar and Carion, Nicolas and Misra, Ishan},
  booktitle={2023 IEEE/CVF Conference on Computer Vision and Pattern Recognition (CVPR)},
  pages={6862--6872},
  year={2023},
  organization={IEEE}
}

@inproceedings{yuksekgonul2023aro,
  title={When and why vision-language models behave like bags-of-words, and what to do about it?},
  author={Yuksekgonul, Mert and Bianchi, Federico and Kalluri, Pratyusha and Jurafsky, Dan and Zou, James},
  journal={arXiv preprint arXiv:2210.01936},
  year={2022}
}

@article{hsieh2023sugarcrepe,
  title={Sugarcrepe: Fixing hackable benchmarks for vision-language compositionality},
  author={Hsieh, Cheng-Yu and Zhang, Jieyu and Ma, Zixian and Kembhavi, Aniruddha and Krishna, Ranjay},
  journal={Advances in neural information processing systems},
  volume={36},
  pages={31096--31116},
  year={2023}
}

@inproceedings{abdin2023kitab,
 title={Kitab: Evaluating llms on constraint satisfaction for information retrieval},
  author={Abdin, Marah I and Gunasekar, Suriya and Chandrasekaran, Varun and Li, Jerry and Yuksekgonul, Mert and Peshawaria, Rahee and Naik, Ranjita and Nushi, Besmira},
  booktitle={International Conference on Learning Representations},
  volume={2024},
  pages={30664--30686},
  year={2024}
}

@inproceedings{malaviya2023quest,
  title={Quest: A retrieval dataset of entity-seeking queries with implicit set operations},
  author={Malaviya, Chaitanya and Shaw, Peter and Chang, Ming-Wei and Lee, Kenton and Toutanova, Kristina},
  booktitle={Proceedings of the 61st Annual Meeting of the Association for Computational Linguistics (Volume 1: Long Papers)},
  pages={14032--14047},
  year={2023}
}

@inproceedings{rahmani2024synthetic,
  title={Synthetic test collections for retrieval evaluation},
  author={Rahmani, Hossein A and Craswell, Nick and Yilmaz, Emine and Mitra, Bhaskar and Campos, Daniel},
  booktitle={Proceedings of the 47th International ACM SIGIR Conference on Research and Development in Information Retrieval},
  pages={2647--2651},
  year={2024}
}

@article{fang2025hidvae,
  title={Hid-vae: Interpretable generative recommendation via hierarchical and disentangled semantic ids},
  author={Fang, Dengzhao and Gao, Jingtong and Zhu, Chengcheng and Li, Yu and Zhao, Xiangyu and Chang, Yi},
  journal={arXiv preprint arXiv:2508.04618},
  year={2025}
}

@misc{khosla2021supervisedcontrastivelearning,
      title={Supervised Contrastive Learning}, 
      author={Prannay Khosla and Piotr Teterwak and Chen Wang and Aaron Sarna and Yonglong Tian and Phillip Isola and Aaron Maschinot and Ce Liu and Dilip Krishnan},
      year={2021},
      eprint={2004.11362},
      archivePrefix={arXiv},
      primaryClass={cs.LG},
      url={https://arxiv.org/abs/2004.11362}, 
}

@misc{param_aggarwal_2019,
	title={Fashion Product Images Dataset},
	url={https://www.kaggle.com/ds/139630},
	DOI={10.34740/KAGGLE/DS/139630},
	publisher={Kaggle},
	author={Param Aggarwal},
	year={2019}
}

@article{Robertson1977TheC1,
  title={The CIE 1976 Color-Difference Formulae},
  author={Alan R. Robertson},
  journal={Color Research and Application},
  year={1977},
  volume={2},
  pages={7-11},
  url={https://api.semanticscholar.org/CorpusID:122447977}
}

@article{reddy2022shopping,
title={Shopping Queries Dataset: A Large-Scale {ESCI} Benchmark for Improving Product Search},
author={Chandan K. Reddy and Lluís Màrquez and Fran Valero and Nikhil Rao and Hugo Zaragoza and Sambaran Bandyopadhyay and Arnab Biswas and Anlu Xing and Karthik Subbian},
year={2022},
eprint={2206.06588},
archivePrefix={arXiv}
}

@book{robertson2009probabilistic,
  title={The probabilistic relevance framework: BM25 and beyond},
  author={Robertson, Stephen and Zaragoza, Hugo},
  year={2009},
  publisher={Now Publishers Inc}
}

@article{qwen3_embedding_2025,
  title   = {Qwen3 Embedding: Advancing Text Embedding and Reranking},
  author  = {Qwen Team},
  journal = {arXiv preprint arXiv:2506.05176},
  year    = {2025},
  url     = {https://arxiv.org/abs/2506.05176}
}

@article{liu2019roberta,
  title={RoBERTa: A Robustly Optimized BERT Pretraining Approach},
  author={Liu, Yinhan and Ott, Myle and Goyal, Naman and Jingfei, Du and Joshi, Mandar and Chen, Danqi and Levy, Omer and Lewis, Mike and Zettlemoyer, Luke and Stoyanov, Veselin},
  journal={arXiv preprint arXiv:1907.11692},
  year={2019}
}

@article{wang2023improving,
  title={Improving Text Embeddings with Large Language Models},
  author={Wang, Liang and Yang, Nan and Huang, Xiaolong and Yang, Linjun and Majumder, Rangan and Wei, Furu},
  journal={arXiv preprint arXiv:2401.00368},
  year={2023}
}

@misc{wu2022smoothedcontrastivelearningunsupervised,
      title={Smoothed Contrastive Learning for Unsupervised Sentence Embedding}, 
      author={Xing Wu and Chaochen Gao and Yipeng Su and Jizhong Han and Zhongyuan Wang and Songlin Hu},
      year={2022},
      eprint={2109.04321},
      archivePrefix={arXiv},
      primaryClass={cs.CL},
      url={https://arxiv.org/abs/2109.04321}, 
}

@inproceedings{ni-etal-2022-sentence,
    title = {Sentence-{T}5: Scalable Sentence Encoders from Pre-trained Text-to-Text Models},
    author = {Ni, Jianmo  and
      Hern{\'a}ndez {\'A}brego, Gustavo  and
      Constant, Noah  and
      Ma, Ji  and
      Hall, Keith  and
      Cer, Daniel  and
      Yang, Yinfei},
    booktitle = {Findings of the Association for Computational Linguistics: ACL 2022},
    month = {may},
    year = {2022},
    address = {Dublin, Ireland},
    publisher = {Association for Computational Linguistics},
    url = {https://aclanthology.org/2022.findings-acl.146/},
    pages = {1864--1874}
}

@misc{SFRAIResearch2024,
  title        = {SFR-Embedding-Mistral: Enhance Text Retrieval with Transfer Learning},
  author       = {Rui Meng and Ye Liu and Shafiq Rayhan Joty and Caiming Xiong and Yingbo Zhou and Semih Yavuz},
  howpublished = {Salesforce AI Research Blog},
  year         = {2024},
  url          = {https://www.salesforce.com/blog/sfr-embedding/}
}

@article{Cronbach_1951, title={Coefficient Alpha and the Internal Structure of Tests}, volume={16}, DOI={10.1007/BF02310555}, number={3}, journal={Psychometrika}, author={Cronbach, Lee J.}, year={1951}, pages={297–334}}

@misc{geminiteam2025geminifamilyhighlycapable,
      title={Gemini: A Family of Highly Capable Multimodal Models}, 
      author={Gemini Team and Rohan Anil and Sebastian Borgeaud and Jean-Baptiste Alayrac and Jiahui Yu and Radu Soricut and Johan Schalkwyk and Andrew M. Dai and Anja Hauth and Katie Millican and David Silver and Melvin Johnson and Ioannis Antonoglou and Julian Schrittwieser and Amelia Glaese and Jilin Chen and Emily Pitler and Timothy Lillicrap and Angeliki Lazaridou and Orhan Firat and James Molloy and Michael Isard and Paul R. Barham and Tom Hennigan and Benjamin Lee and Fabio Viola and Malcolm Reynolds and Yuanzhong Xu and Ryan Doherty and Eli Collins and Clemens Meyer and Eliza Rutherford and Erica Moreira and Kareem Ayoub and Megha Goel and Jack Krawczyk and Cosmo Du and Ed Chi and Heng-Tze Cheng and Eric Ni and Purvi Shah and Patrick Kane and Betty Chan and Manaal Faruqui and Aliaksei Severyn and Hanzhao Lin and YaGuang Li and Yong Cheng and Abe Ittycheriah and Mahdis Mahdieh and Mia Chen and Pei Sun and Dustin Tran and Sumit Bagri and Balaji Lakshminarayanan and Jeremiah Liu and Andras Orban and Fabian Güra and Hao Zhou and Xinying Song and Aurelien Boffy and Harish Ganapathy and Steven Zheng and HyunJeong Choe and Ágoston Weisz and Tao Zhu and Yifeng Lu and Siddharth Gopal and Jarrod Kahn and Maciej Kula and Jeff Pitman and Rushin Shah and Emanuel Taropa and Majd Al Merey and Martin Baeuml and Zhifeng Chen and Laurent El Shafey and Yujing Zhang and Olcan Sercinoglu and George Tucker and Enrique Piqueras and Maxim Krikun and Iain Barr and Nikolay Savinov and Ivo Danihelka and Becca Roelofs and Anaïs White and Anders Andreassen and Tamara von Glehn and Lakshman Yagati and Mehran Kazemi and Lucas Gonzalez and Misha Khalman and Jakub Sygnowski and Alexandre Frechette and Charlotte Smith and Laura Culp and Lev Proleev and Yi Luan and Xi Chen and James Lottes and Nathan Schucher and Federico Lebron and Alban Rrustemi and Natalie Clay and Phil Crone and Tomas Kocisky and Jeffrey Zhao and Bartek Perz and Dian Yu and Heidi Howard and Adam Bloniarz and Jack W. Rae and Han Lu and Laurent Sifre and Marcello Maggioni and Fred Alcober and Dan Garrette and Megan Barnes and Shantanu Thakoor and Jacob Austin and Gabriel Barth-Maron and William Wong and Rishabh Joshi and Rahma Chaabouni and Deeni Fatiha and Arun Ahuja and Gaurav Singh Tomar and Evan Senter and Martin Chadwick and Ilya Kornakov and Nithya Attaluri and Iñaki Iturrate and Ruibo Liu and Yunxuan Li and Sarah Cogan and Jeremy Chen and Chao Jia and Chenjie Gu and Qiao Zhang and Jordan Grimstad and Ale Jakse Hartman and Xavier Garcia and Thanumalayan Sankaranarayana Pillai and Jacob Devlin and Michael Laskin and Diego de Las Casas and Dasha Valter and Connie Tao and Lorenzo Blanco and Adrià Puigdomènech Badia and David Reitter and Mianna Chen and Jenny Brennan and Clara Rivera and Sergey Brin and Shariq Iqbal and Gabriela Surita and Jane Labanowski and Abhi Rao and Stephanie Winkler and Emilio Parisotto and Yiming Gu and Kate Olszewska and Ravi Addanki and Antoine Miech and Annie Louis and Denis Teplyashin and Geoff Brown and Elliot Catt and Jan Balaguer and Jackie Xiang and Pidong Wang and Zoe Ashwood and Anton Briukhov and Albert Webson and Sanjay Ganapathy and Smit Sanghavi and Ajay Kannan and Ming-Wei Chang and Axel Stjerngren and Josip Djolonga and Yuting Sun and Ankur Bapna and Matthew Aitchison and Pedram Pejman and Henryk Michalewski and Tianhe Yu and Cindy Wang and Juliette Love and Junwhan Ahn and Dawn Bloxwich and Kehang Han and Peter Humphreys and Thibault Sellam and James Bradbury and Varun Godbole and Sina Samangooei and Bogdan Damoc and Alex Kaskasoli and Sébastien M. R. Arnold and Vijay Vasudevan and Shubham Agrawal and Jason Riesa and Dmitry Lepikhin and Richard Tanburn and Srivatsan Srinivasan and Hyeontaek Lim and Sarah Hodkinson and Pranav Shyam and Johan Ferret and Steven Hand and Ankush Garg and Tom Le Paine and Jian Li and Yujia Li and Minh Giang and Alexander Neitz and Zaheer Abbas and Sarah York and Machel Reid and Elizabeth Cole and Aakanksha Chowdhery and Dipanjan Das and Dominika Rogozińska and Vitaliy Nikolaev and Pablo Sprechmann and Zachary Nado and Lukas Zilka and Flavien Prost and Luheng He and Marianne Monteiro and Gaurav Mishra and Chris Welty and Josh Newlan and Dawei Jia and Miltiadis Allamanis and Clara Huiyi Hu and Raoul de Liedekerke and Justin Gilmer and Carl Saroufim and Shruti Rijhwani and Shaobo Hou and Disha Shrivastava and Anirudh Baddepudi and Alex Goldin and Adnan Ozturel and Albin Cassirer and Yunhan Xu and Daniel Sohn and Devendra Sachan and Reinald Kim Amplayo and Craig Swanson and Dessie Petrova and Shashi Narayan and Arthur Guez and Siddhartha Brahma and Jessica Landon and Miteyan Patel and Ruizhe Zhao and Kevin Villela and Luyu Wang and Wenhao Jia and Matthew Rahtz and Mai Giménez and Legg Yeung and James Keeling and Petko Georgiev and Diana Mincu and Boxi Wu and Salem Haykal and Rachel Saputro and Kiran Vodrahalli and James Qin and Zeynep Cankara and Abhanshu Sharma and Nick Fernando and Will Hawkins and Behnam Neyshabur and Solomon Kim and Adrian Hutter and Priyanka Agrawal and Alex Castro-Ros and George van den Driessche and Tao Wang and Fan Yang and Shuo-yiin Chang and Paul Komarek and Ross McIlroy and Mario Lučić and Guodong Zhang and Wael Farhan and Michael Sharman and Paul Natsev and Paul Michel and Yamini Bansal and Siyuan Qiao and Kris Cao and Siamak Shakeri and Christina Butterfield and Justin Chung and Paul Kishan Rubenstein and Shivani Agrawal and Arthur Mensch and Kedar Soparkar and Karel Lenc and Timothy Chung and Aedan Pope and Loren Maggiore and Jackie Kay and Priya Jhakra and Shibo Wang and Joshua Maynez and Mary Phuong and Taylor Tobin and Andrea Tacchetti and Maja Trebacz and Kevin Robinson and Yash Katariya and Sebastian Riedel and Paige Bailey and Kefan Xiao and Nimesh Ghelani and Lora Aroyo and Ambrose Slone and Neil Houlsby and Xuehan Xiong and Zhen Yang and Elena Gribovskaya and Jonas Adler and Mateo Wirth and Lisa Lee and Music Li and Thais Kagohara and Jay Pavagadhi and Sophie Bridgers and Anna Bortsova and Sanjay Ghemawat and Zafarali Ahmed and Tianqi Liu and Richard Powell and Vijay Bolina and Mariko Iinuma and Polina Zablotskaia and James Besley and Da-Woon Chung and Timothy Dozat and Ramona Comanescu and Xiance Si and Jeremy Greer and Guolong Su and Martin Polacek and Raphaël Lopez Kaufman and Simon Tokumine and Hexiang Hu and Elena Buchatskaya and Yingjie Miao and Mohamed Elhawaty and Aditya Siddhant and Nenad Tomasev and Jinwei Xing and Christina Greer and Helen Miller and Shereen Ashraf and Aurko Roy and Zizhao Zhang and Ada Ma and Angelos Filos and Milos Besta and Rory Blevins and Ted Klimenko and Chih-Kuan Yeh and Soravit Changpinyo and Jiaqi Mu and Oscar Chang and Mantas Pajarskas and Carrie Muir and Vered Cohen and Charline Le Lan and Krishna Haridasan and Amit Marathe and Steven Hansen and Sholto Douglas and Rajkumar Samuel and Mingqiu Wang and Sophia Austin and Chang Lan and Jiepu Jiang and Justin Chiu and Jaime Alonso Lorenzo and Lars Lowe Sjösund and Sébastien Cevey and Zach Gleicher and Thi Avrahami and Anudhyan Boral and Hansa Srinivasan and Vittorio Selo and Rhys May and Konstantinos Aisopos and Léonard Hussenot and Livio Baldini Soares and Kate Baumli and Michael B. Chang and Adrià Recasens and Ben Caine and Alexander Pritzel and Filip Pavetic and Fabio Pardo and Anita Gergely and Justin Frye and Vinay Ramasesh and Dan Horgan and Kartikeya Badola and Nora Kassner and Subhrajit Roy and Ethan Dyer and Víctor Campos Campos and Alex Tomala and Yunhao Tang and Dalia El Badawy and Elspeth White and Basil Mustafa and Oran Lang and Abhishek Jindal and Sharad Vikram and Zhitao Gong and Sergi Caelles and Ross Hemsley and Gregory Thornton and Fangxiaoyu Feng and Wojciech Stokowiec and Ce Zheng and Phoebe Thacker and Çağlar Ünlü and Zhishuai Zhang and Mohammad Saleh and James Svensson and Max Bileschi and Piyush Patil and Ankesh Anand and Roman Ring and Katerina Tsihlas and Arpi Vezer and Marco Selvi and Toby Shevlane and Mikel Rodriguez and Tom Kwiatkowski and Samira Daruki and Keran Rong and Allan Dafoe and Nicholas FitzGerald and Keren Gu-Lemberg and Mina Khan and Lisa Anne Hendricks and Marie Pellat and Vladimir Feinberg and James Cobon-Kerr and Tara Sainath and Maribeth Rauh and Sayed Hadi Hashemi and Richard Ives and Yana Hasson and Eric Noland and Yuan Cao and Nathan Byrd and Le Hou and Qingze Wang and Thibault Sottiaux and Michela Paganini and Jean-Baptiste Lespiau and Alexandre Moufarek and Samer Hassan and Kaushik Shivakumar and Joost van Amersfoort and Amol Mandhane and Pratik Joshi and Anirudh Goyal and Matthew Tung and Andrew Brock and Hannah Sheahan and Vedant Misra and Cheng Li and Nemanja Rakićević and Mostafa Dehghani and Fangyu Liu and Sid Mittal and Junhyuk Oh and Seb Noury and Eren Sezener and Fantine Huot and Matthew Lamm and Nicola De Cao and Charlie Chen and Sidharth Mudgal and Romina Stella and Kevin Brooks and Gautam Vasudevan and Chenxi Liu and Mainak Chain and Nivedita Melinkeri and Aaron Cohen and Venus Wang and Kristie Seymore and Sergey Zubkov and Rahul Goel and Summer Yue and Sai Krishnakumaran and Brian Albert and Nate Hurley and Motoki Sano and Anhad Mohananey and Jonah Joughin and Egor Filonov and Tomasz Kępa and Yomna Eldawy and Jiawern Lim and Rahul Rishi and Shirin Badiezadegan and Taylor Bos and Jerry Chang and Sanil Jain and Sri Gayatri Sundara Padmanabhan and Subha Puttagunta and Kalpesh Krishna and Leslie Baker and Norbert Kalb and Vamsi Bedapudi and Adam Kurzrok and Shuntong Lei and Anthony Yu and Oren Litvin and Xiang Zhou and Zhichun Wu and Sam Sobell and Andrea Siciliano and Alan Papir and Robby Neale and Jonas Bragagnolo and Tej Toor and Tina Chen and Valentin Anklin and Feiran Wang and Richie Feng and Milad Gholami and Kevin Ling and Lijuan Liu and Jules Walter and Hamid Moghaddam and Arun Kishore and Jakub Adamek and Tyler Mercado and Jonathan Mallinson and Siddhinita Wandekar and Stephen Cagle and Eran Ofek and Guillermo Garrido and Clemens Lombriser and Maksim Mukha and Botu Sun and Hafeezul Rahman Mohammad and Josip Matak and Yadi Qian and Vikas Peswani and Pawel Janus and Quan Yuan and Leif Schelin and Oana David and Ankur Garg and Yifan He and Oleksii Duzhyi and Anton Älgmyr and Timothée Lottaz and Qi Li and Vikas Yadav and Luyao Xu and Alex Chinien and Rakesh Shivanna and Aleksandr Chuklin and Josie Li and Carrie Spadine and Travis Wolfe and Kareem Mohamed and Subhabrata Das and Zihang Dai and Kyle He and Daniel von Dincklage and Shyam Upadhyay and Akanksha Maurya and Luyan Chi and Sebastian Krause and Khalid Salama and Pam G Rabinovitch and Pavan Kumar Reddy M and Aarush Selvan and Mikhail Dektiarev and Golnaz Ghiasi and Erdem Guven and Himanshu Gupta and Boyi Liu and Deepak Sharma and Idan Heimlich Shtacher and Shachi Paul and Oscar Akerlund and François-Xavier Aubet and Terry Huang and Chen Zhu and Eric Zhu and Elico Teixeira and Matthew Fritze and Francesco Bertolini and Liana-Eleonora Marinescu and Martin Bölle and Dominik Paulus and Khyatti Gupta and Tejasi Latkar and Max Chang and Jason Sanders and Roopa Wilson and Xuewei Wu and Yi-Xuan Tan and Lam Nguyen Thiet and Tulsee Doshi and Sid Lall and Swaroop Mishra and Wanming Chen and Thang Luong and Seth Benjamin and Jasmine Lee and Ewa Andrejczuk and Dominik Rabiej and Vipul Ranjan and Krzysztof Styrc and Pengcheng Yin and Jon Simon and Malcolm Rose Harriott and Mudit Bansal and Alexei Robsky and Geoff Bacon and David Greene and Daniil Mirylenka and Chen Zhou and Obaid Sarvana and Abhimanyu Goyal and Samuel Andermatt and Patrick Siegler and Ben Horn and Assaf Israel and Francesco Pongetti and Chih-Wei "Louis" Chen and Marco Selvatici and Pedro Silva and Kathie Wang and Jackson Tolins and Kelvin Guu and Roey Yogev and Xiaochen Cai and Alessandro Agostini and Maulik Shah and Hung Nguyen and Noah Ó Donnaile and Sébastien Pereira and Linda Friso and Adam Stambler and Adam Kurzrok and Chenkai Kuang and Yan Romanikhin and Mark Geller and ZJ Yan and Kane Jang and Cheng-Chun Lee and Wojciech Fica and Eric Malmi and Qijun Tan and Dan Banica and Daniel Balle and Ryan Pham and Yanping Huang and Diana Avram and Hongzhi Shi and Jasjot Singh and Chris Hidey and Niharika Ahuja and Pranab Saxena and Dan Dooley and Srividya Pranavi Potharaju and Eileen O'Neill and Anand Gokulchandran and Ryan Foley and Kai Zhao and Mike Dusenberry and Yuan Liu and Pulkit Mehta and Ragha Kotikalapudi and Chalence Safranek-Shrader and Andrew Goodman and Joshua Kessinger and Eran Globen and Prateek Kolhar and Chris Gorgolewski and Ali Ibrahim and Yang Song and Ali Eichenbaum and Thomas Brovelli and Sahitya Potluri and Preethi Lahoti and Cip Baetu and Ali Ghorbani and Charles Chen and Andy Crawford and Shalini Pal and Mukund Sridhar and Petru Gurita and Asier Mujika and Igor Petrovski and Pierre-Louis Cedoz and Chenmei Li and Shiyuan Chen and Niccolò Dal Santo and Siddharth Goyal and Jitesh Punjabi and Karthik Kappaganthu and Chester Kwak and Pallavi LV and Sarmishta Velury and Himadri Choudhury and Jamie Hall and Premal Shah and Ricardo Figueira and Matt Thomas and Minjie Lu and Ting Zhou and Chintu Kumar and Thomas Jurdi and Sharat Chikkerur and Yenai Ma and Adams Yu and Soo Kwak and Victor Ähdel and Sujeevan Rajayogam and Travis Choma and Fei Liu and Aditya Barua and Colin Ji and Ji Ho Park and Vincent Hellendoorn and Alex Bailey and Taylan Bilal and Huanjie Zhou and Mehrdad Khatir and Charles Sutton and Wojciech Rzadkowski and Fiona Macintosh and Roopali Vij and Konstantin Shagin and Paul Medina and Chen Liang and Jinjing Zhou and Pararth Shah and Yingying Bi and Attila Dankovics and Shipra Banga and Sabine Lehmann and Marissa Bredesen and Zifan Lin and John Eric Hoffmann and Jonathan Lai and Raynald Chung and Kai Yang and Nihal Balani and Arthur Bražinskas and Andrei Sozanschi and Matthew Hayes and Héctor Fernández Alcalde and Peter Makarov and Will Chen and Antonio Stella and Liselotte Snijders and Michael Mandl and Ante Kärrman and Paweł Nowak and Xinyi Wu and Alex Dyck and Krishnan Vaidyanathan and Raghavender R and Jessica Mallet and Mitch Rudominer and Eric Johnston and Sushil Mittal and Akhil Udathu and Janara Christensen and Vishal Verma and Zach Irving and Andreas Santucci and Gamaleldin Elsayed and Elnaz Davoodi and Marin Georgiev and Ian Tenney and Nan Hua and Geoffrey Cideron and Edouard Leurent and Mahmoud Alnahlawi and Ionut Georgescu and Nan Wei and Ivy Zheng and Dylan Scandinaro and Heinrich Jiang and Jasper Snoek and Mukund Sundararajan and Xuezhi Wang and Zack Ontiveros and Itay Karo and Jeremy Cole and Vinu Rajashekhar and Lara Tumeh and Eyal Ben-David and Rishub Jain and Jonathan Uesato and Romina Datta and Oskar Bunyan and Shimu Wu and John Zhang and Piotr Stanczyk and Ye Zhang and David Steiner and Subhajit Naskar and Michael Azzam and Matthew Johnson and Adam Paszke and Chung-Cheng Chiu and Jaume Sanchez Elias and Afroz Mohiuddin and Faizan Muhammad and Jin Miao and Andrew Lee and Nino Vieillard and Jane Park and Jiageng Zhang and Jeff Stanway and Drew Garmon and Abhijit Karmarkar and Zhe Dong and Jong Lee and Aviral Kumar and Luowei Zhou and Jonathan Evens and William Isaac and Geoffrey Irving and Edward Loper and Michael Fink and Isha Arkatkar and Nanxin Chen and Izhak Shafran and Ivan Petrychenko and Zhe Chen and Johnson Jia and Anselm Levskaya and Zhenkai Zhu and Peter Grabowski and Yu Mao and Alberto Magni and Kaisheng Yao and Javier Snaider and Norman Casagrande and Evan Palmer and Paul Suganthan and Alfonso Castaño and Irene Giannoumis and Wooyeol Kim and Mikołaj Rybiński and Ashwin Sreevatsa and Jennifer Prendki and David Soergel and Adrian Goedeckemeyer and Willi Gierke and Mohsen Jafari and Meenu Gaba and Jeremy Wiesner and Diana Gage Wright and Yawen Wei and Harsha Vashisht and Yana Kulizhskaya and Jay Hoover and Maigo Le and Lu Li and Chimezie Iwuanyanwu and Lu Liu and Kevin Ramirez and Andrey Khorlin and Albert Cui and Tian LIN and Marcus Wu and Ricardo Aguilar and Keith Pallo and Abhishek Chakladar and Ginger Perng and Elena Allica Abellan and Mingyang Zhang and Ishita Dasgupta and Nate Kushman and Ivo Penchev and Alena Repina and Xihui Wu and Tom van der Weide and Priya Ponnapalli and Caroline Kaplan and Jiri Simsa and Shuangfeng Li and Olivier Dousse and Fan Yang and Jeff Piper and Nathan Ie and Rama Pasumarthi and Nathan Lintz and Anitha Vijayakumar and Daniel Andor and Pedro Valenzuela and Minnie Lui and Cosmin Paduraru and Daiyi Peng and Katherine Lee and Shuyuan Zhang and Somer Greene and Duc Dung Nguyen and Paula Kurylowicz and Cassidy Hardin and Lucas Dixon and Lili Janzer and Kiam Choo and Ziqiang Feng and Biao Zhang and Achintya Singhal and Dayou Du and Dan McKinnon and Natasha Antropova and Tolga Bolukbasi and Orgad Keller and David Reid and Daniel Finchelstein and Maria Abi Raad and Remi Crocker and Peter Hawkins and Robert Dadashi and Colin Gaffney and Ken Franko and Anna Bulanova and Rémi Leblond and Shirley Chung and Harry Askham and Luis C. Cobo and Kelvin Xu and Felix Fischer and Jun Xu and Christina Sorokin and Chris Alberti and Chu-Cheng Lin and Colin Evans and Alek Dimitriev and Hannah Forbes and Dylan Banarse and Zora Tung and Mark Omernick and Colton Bishop and Rachel Sterneck and Rohan Jain and Jiawei Xia and Ehsan Amid and Francesco Piccinno and Xingyu Wang and Praseem Banzal and Daniel J. Mankowitz and Alex Polozov and Victoria Krakovna and Sasha Brown and MohammadHossein Bateni and Dennis Duan and Vlad Firoiu and Meghana Thotakuri and Tom Natan and Matthieu Geist and Ser tan Girgin and Hui Li and Jiayu Ye and Ofir Roval and Reiko Tojo and Michael Kwong and James Lee-Thorp and Christopher Yew and Danila Sinopalnikov and Sabela Ramos and John Mellor and Abhishek Sharma and Kathy Wu and David Miller and Nicolas Sonnerat and Denis Vnukov and Rory Greig and Jennifer Beattie and Emily Caveness and Libin Bai and Julian Eisenschlos and Alex Korchemniy and Tomy Tsai and Mimi Jasarevic and Weize Kong and Phuong Dao and Zeyu Zheng and Frederick Liu and Fan Yang and Rui Zhu and Tian Huey Teh and Jason Sanmiya and Evgeny Gladchenko and Nejc Trdin and Daniel Toyama and Evan Rosen and Sasan Tavakkol and Linting Xue and Chen Elkind and Oliver Woodman and John Carpenter and George Papamakarios and Rupert Kemp and Sushant Kafle and Tanya Grunina and Rishika Sinha and Alice Talbert and Diane Wu and Denese Owusu-Afriyie and Cosmo Du and Chloe Thornton and Jordi Pont-Tuset and Pradyumna Narayana and Jing Li and Saaber Fatehi and John Wieting and Omar Ajmeri and Benigno Uria and Yeongil Ko and Laura Knight and Amélie Héliou and Ning Niu and Shane Gu and Chenxi Pang and Yeqing Li and Nir Levine and Ariel Stolovich and Rebeca Santamaria-Fernandez and Sonam Goenka and Wenny Yustalim and Robin Strudel and Ali Elqursh and Charlie Deck and Hyo Lee and Zonglin Li and Kyle Levin and Raphael Hoffmann and Dan Holtmann-Rice and Olivier Bachem and Sho Arora and Christy Koh and Soheil Hassas Yeganeh and Siim Põder and Mukarram Tariq and Yanhua Sun and Lucian Ionita and Mojtaba Seyedhosseini and Pouya Tafti and Zhiyu Liu and Anmol Gulati and Jasmine Liu and Xinyu Ye and Bart Chrzaszcz and Lily Wang and Nikhil Sethi and Tianrun Li and Ben Brown and Shreya Singh and Wei Fan and Aaron Parisi and Joe Stanton and Vinod Koverkathu and Christopher A. Choquette-Choo and Yunjie Li and TJ Lu and Abe Ittycheriah and Prakash Shroff and Mani Varadarajan and Sanaz Bahargam and Rob Willoughby and David Gaddy and Guillaume Desjardins and Marco Cornero and Brona Robenek and Bhavishya Mittal and Ben Albrecht and Ashish Shenoy and Fedor Moiseev and Henrik Jacobsson and Alireza Ghaffarkhah and Morgane Rivière and Alanna Walton and Clément Crepy and Alicia Parrish and Zongwei Zhou and Clement Farabet and Carey Radebaugh and Praveen Srinivasan and Claudia van der Salm and Andreas Fidjeland and Salvatore Scellato and Eri Latorre-Chimoto and Hanna Klimczak-Plucińska and David Bridson and Dario de Cesare and Tom Hudson and Piermaria Mendolicchio and Lexi Walker and Alex Morris and Matthew Mauger and Alexey Guseynov and Alison Reid and Seth Odoom and Lucia Loher and Victor Cotruta and Madhavi Yenugula and Dominik Grewe and Anastasia Petrushkina and Tom Duerig and Antonio Sanchez and Steve Yadlowsky and Amy Shen and Amir Globerson and Lynette Webb and Sahil Dua and Dong Li and Surya Bhupatiraju and Dan Hurt and Haroon Qureshi and Ananth Agarwal and Tomer Shani and Matan Eyal and Anuj Khare and Shreyas Rammohan Belle and Lei Wang and Chetan Tekur and Mihir Sanjay Kale and Jinliang Wei and Ruoxin Sang and Brennan Saeta and Tyler Liechty and Yi Sun and Yao Zhao and Stephan Lee and Pandu Nayak and Doug Fritz and Manish Reddy Vuyyuru and John Aslanides and Nidhi Vyas and Martin Wicke and Xiao Ma and Evgenii Eltyshev and Nina Martin and Hardie Cate and James Manyika and Keyvan Amiri and Yelin Kim and Xi Xiong and Kai Kang and Florian Luisier and Nilesh Tripuraneni and David Madras and Mandy Guo and Austin Waters and Oliver Wang and Joshua Ainslie and Jason Baldridge and Han Zhang and Garima Pruthi and Jakob Bauer and Feng Yang and Riham Mansour and Jason Gelman and Yang Xu and George Polovets and Ji Liu and Honglong Cai and Warren Chen and XiangHai Sheng and Emily Xue and Sherjil Ozair and Christof Angermueller and Xiaowei Li and Anoop Sinha and Weiren Wang and Julia Wiesinger and Emmanouil Koukoumidis and Yuan Tian and Anand Iyer and Madhu Gurumurthy and Mark Goldenson and Parashar Shah and MK Blake and Hongkun Yu and Anthony Urbanowicz and Jennimaria Palomaki and Chrisantha Fernando and Ken Durden and Harsh Mehta and Nikola Momchev and Elahe Rahimtoroghi and Maria Georgaki and Amit Raul and Sebastian Ruder and Morgan Redshaw and Jinhyuk Lee and Denny Zhou and Komal Jalan and Dinghua Li and Blake Hechtman and Parker Schuh and Milad Nasr and Kieran Milan and Vladimir Mikulik and Juliana Franco and Tim Green and Nam Nguyen and Joe Kelley and Aroma Mahendru and Andrea Hu and Joshua Howland and Ben Vargas and Jeffrey Hui and Kshitij Bansal and Vikram Rao and Rakesh Ghiya and Emma Wang and Ke Ye and Jean Michel Sarr and Melanie Moranski Preston and Madeleine Elish and Steve Li and Aakash Kaku and Jigar Gupta and Ice Pasupat and Da-Cheng Juan and Milan Someswar and Tejvi M. and Xinyun Chen and Aida Amini and Alex Fabrikant and Eric Chu and Xuanyi Dong and Amruta Muthal and Senaka Buthpitiya and Sarthak Jauhari and Nan Hua and Urvashi Khandelwal and Ayal Hitron and Jie Ren and Larissa Rinaldi and Shahar Drath and Avigail Dabush and Nan-Jiang Jiang and Harshal Godhia and Uli Sachs and Anthony Chen and Yicheng Fan and Hagai Taitelbaum and Hila Noga and Zhuyun Dai and James Wang and Chen Liang and Jenny Hamer and Chun-Sung Ferng and Chenel Elkind and Aviel Atias and Paulina Lee and Vít Listík and Mathias Carlen and Jan van de Kerkhof and Marcin Pikus and Krunoslav Zaher and Paul Müller and Sasha Zykova and Richard Stefanec and Vitaly Gatsko and Christoph Hirnschall and Ashwin Sethi and Xingyu Federico Xu and Chetan Ahuja and Beth Tsai and Anca Stefanoiu and Bo Feng and Keshav Dhandhania and Manish Katyal and Akshay Gupta and Atharva Parulekar and Divya Pitta and Jing Zhao and Vivaan Bhatia and Yashodha Bhavnani and Omar Alhadlaq and Xiaolin Li and Peter Danenberg and Dennis Tu and Alex Pine and Vera Filippova and Abhipso Ghosh and Ben Limonchik and Bhargava Urala and Chaitanya Krishna Lanka and Derik Clive and Yi Sun and Edward Li and Hao Wu and Kevin Hongtongsak and Ianna Li and Kalind Thakkar and Kuanysh Omarov and Kushal Majmundar and Michael Alverson and Michael Kucharski and Mohak Patel and Mudit Jain and Maksim Zabelin and Paolo Pelagatti and Rohan Kohli and Saurabh Kumar and Joseph Kim and Swetha Sankar and Vineet Shah and Lakshmi Ramachandruni and Xiangkai Zeng and Ben Bariach and Laura Weidinger and Tu Vu and Alek Andreev and Antoine He and Kevin Hui and Sheleem Kashem and Amar Subramanya and Sissie Hsiao and Demis Hassabis and Koray Kavukcuoglu and Adam Sadovsky and Quoc Le and Trevor Strohman and Yonghui Wu and Slav Petrov and Jeffrey Dean and Oriol Vinyals},
      year={2025},
      eprint={2312.11805},
      archivePrefix={arXiv},
      primaryClass={cs.CL},
      url={https://arxiv.org/abs/2312.11805}, 
}

@misc{formal2021spladev2sparselexical,
      title={SPLADE v2: Sparse Lexical and Expansion Model for Information Retrieval}, 
      author={Thibault Formal and Carlos Lassance and Benjamin Piwowarski and Stéphane Clinchant},
      year={2021},
      eprint={2109.10086},
      archivePrefix={arXiv},
      primaryClass={cs.IR},
      url={https://arxiv.org/abs/2109.10086}, 
}

@misc{izacard2022unsuperviseddenseinformationretrieval,
      title={Unsupervised Dense Information Retrieval with Contrastive Learning}, 
      author={Gautier Izacard and Mathilde Caron and Lucas Hosseini and Sebastian Riedel and Piotr Bojanowski and Armand Joulin and Edouard Grave},
      year={2022},
      eprint={2112.09118},
      archivePrefix={arXiv},
      primaryClass={cs.IR},
      url={https://arxiv.org/abs/2112.09118}, 
}

@misc{santhanam2022colbertv2effectiveefficientretrieval,
      title={ColBERTv2: Effective and Efficient Retrieval via Lightweight Late Interaction}, 
      author={Keshav Santhanam and Omar Khattab and Jon Saad-Falcon and Christopher Potts and Matei Zaharia},
      year={2022},
      eprint={2112.01488},
      archivePrefix={arXiv},
      primaryClass={cs.IR},
      url={https://arxiv.org/abs/2112.01488}, 
}

@inproceedings{warner-etal-2025-smarter,
    title = "Smarter, Better, Faster, Longer: A Modern Bidirectional Encoder for Fast, Memory Efficient, and Long Context Finetuning and Inference",
    author = {Warner, Benjamin  and
      Chaffin, Antoine  and
      Clavi{\'e}, Benjamin  and
      Weller, Orion  and
      Hallstr{\"o}m, Oskar  and
      Taghadouini, Said  and
      Gallagher, Alexis  and
      Biswas, Raja  and
      Ladhak, Faisal  and
      Aarsen, Tom  and
      Adams, Griffin Thomas  and
      Howard, Jeremy  and
      Poli, Iacopo},
    editor = "Che, Wanxiang  and
      Nabende, Joyce  and
      Shutova, Ekaterina  and
      Pilehvar, Mohammad Taher",
    booktitle = "Proceedings of the 63rd Annual Meeting of the Association for Computational Linguistics (Volume 1: Long Papers)",
    month = jul,
    year = "2025",
    address = "Vienna, Austria",
    publisher = "Association for Computational Linguistics",
    url = "https://aclanthology.org/2025.acl-long.127/",
    doi = "10.18653/v1/2025.acl-long.127",
    pages = "2526--2547",
    ISBN = "979-8-89176-251-0"
}

@misc{moderncestsb2025,
  author = {Miller, D. Lee},
  title = {ModernCE STS: An STS cross encoder model},
  year = {2025},
  publisher = {Hugging Face Hub},
  url = {https://huggingface.co/dleemiller/ModernCE-large-sts},
}

@misc{vanwieringen2026lecturenotesridgeregression,
      title={Lecture notes on ridge regression}, 
      author={Wessel N. van Wieringen},
      year={2026},
      eprint={1509.09169},
      archivePrefix={arXiv},
      primaryClass={stat.ME},
      url={https://arxiv.org/abs/1509.09169}, 
}

@article{marquardt1970generalized,
  title={Generalized inverses, ridge regression, biased linear estimation, and nonlinear estimation},
  author={Marquardt, Donald W.},
  journal={Technometrics},
  volume={12},
  number={3},
  pages={591--612},
  year={1970},
  publisher={Taylor \& Francis}
}

@inproceedings{kamalloo2024resources,
  author    = {Kamalloo, Ehsan and Thakur, Nandan and Lassance, Carlos and Ma, Xueguang and Yang, Jheng-Hong and Lin, Jimmy},
  title     = {Resources for Brewing BEIR: Reproducible Reference Models and Statistical Analyses},
  booktitle = {Proceedings of the 47th International ACM SIGIR Conference on Research and Development in Information Retrieval},
  series    = {SIGIR '24},
  pages     = {1431--1440},
  year      = {2024},
  publisher = {Association for Computing Machinery},
  address   = {New York, NY, USA},
  doi       = {10.1145/3626772.3657862},
  url       = {https://doi.org/10.1145/3626772.3657862}
}
}

\end{document}